\documentclass[a4paper,11pt]{article}
\pdfoutput=1 

\usepackage{jheppub} 
\usepackage{xspace}
\usepackage{siunitx}
\usepackage[T1]{fontenc} 
\usepackage{todonotes}
\usepackage{booktabs}
\usepackage{multirow}

\usepackage[section]{placeins}

\title{\boldmath Charged Particle Tracking with Machine Learning on FPGAs}

\newcommand*{\pt}{\ensuremath{p_{\text{T}}}\xspace}
\newcommand*{\electronvolt}{\text{e\kern-0.1em V}}

\newcommand*{\GeV}{\ensuremath{\text{G\electronvolt}}}

\author[a]{H. Abidi,}
\author[b]{A. Boveia,}
\author[a]{V. Cavaliere,}
\author[c]{D. Furletov,}
\author[b]{A. Gekow,}
\author[a]{C. W. Kalderon,}
\author[a]{S. Yoo}

\affiliation[a]{Brookhaven National Laboratory}
\affiliation[b]{Ohio State University}
\affiliation[c]{William and Mary}

\emailAdd{sabidi@bnl.gov}
\emailAdd{antonio.boveia@cern.ch}
\emailAdd{vcavaliere@bnl.gov}
\emailAdd{denis.furletov@gmail.com}
\emailAdd{gekow.1@osu.edu}
\emailAdd{william.kalderon@cern.ch}
\emailAdd{sjyoo@bnl.gov}

\abstract{
The determination of charged particle trajectories (tracking) in collisions at the CERN Large Hadron Collider (LHC) is one of the most important aspects for event reconstruction at hadron colliders.   This is especially true in the high conditions expected during the future high-luminosity phase of the LHC (HL-LHC) where the number of interactions per beam crossing will
increase by a factor of five. Deep learning algorithms have been successfully applied to this task for offline applications.  However, their study in hardware-based trigger applications has been limited . In this paper, we study different algorithms for two different steps of tracking and show that such algorithms can be run on field-programmable gate arrays (FPGAs). }

\begin{document} 
\maketitle
\flushbottom

\section{Introduction}
\label{sec:intro}

Charged-particle tracking is one of the most important aspects for event reconstruction at hadron colliders such as the LHC~\cite{Evans:2008}. Silicon-based, high-granularity tracking sensors detect ionization charge deposited by particles as they propagate through the magnetic field of the detector. Tracking algorithms use this information to measure the curvature of particle trajectories and thus deduce the particles’ charge and momentum. Precise tracking is an indispensable tool for any collider experiment. Efficient identification of electrons and muons, based on tracking, is necessary to separate new phenomena from the overwhelming background. The capability to reconstruct detached vertices is an essential tool, for example, for identifying b-jets in all cases when the new particles have a preferential decay to heavy quarks, like the Higgs boson. 
In particular track reconstruction in the real-time trigger systems is critical to fully exploit the physics capabilities of hadron-collider experiments such as ATLAS~\cite{ATLAS:2008xda}, enabling an early rejection of background and more signal-like events written to tape.
The hardware (software) trigger system at current LHC experiments must make physics decisions within a few microseconds (seconds). Events that do not have a positive trigger decision are not written on tape and are lost. Field Programmable Gate Arrays (FPGAs) are the preferred technology for these high-throughput low-latency situations, as current CPU devices are simply too slow to meet experimental requirements. The flexibility offered by FPGAs is particularly important as they can be updated to meet new experimental needs and can incorporate the latest technological advances. 
This flexibility will be even more important at the next generation colliders, such as the High-Luminosity LHC (HL-LHC ~\cite{Apollinari:2017lan}) or FCC-hh~\cite{FCC-hh} which will bring considerable performance challenges to particle tracking. The average number of interactions per beam crossing will increase by a factor of five (20 for future colliders) compared to current LHC conditions and result in a similar increase in the number of charged particles and hence detector occupancy.

Traditional tracking algorithms scale quadratically or worse with detector occupancy. Under reasonable assumptions on the evolution of CPU performance and cost tracking algorithms will need to become one order of magnitude faster and run in parallel on one order of magnitude more processing units (cores or threads). 
Software based algorithms are therefore not suitable for real-time triggering in low-latency environments. 
The targeted latency can be achieved by implementing advanced machine learning algorithms for track reconstruction in FPGAs to perform tracking at high data rates and low latency.
In particular Deep neural networks (DNNs), such as convolutional neural networks (CNNs)~\cite{CNNS}, recurrent neural networks~\cite{RNNS}, have the stunning capability of analyzing large amount of data and/or making decisions by automatically learning features from data rather than a handcrafted approach. DNNs have demonstrated excellent performance in areas such as computer vision, natural language processing, and robotics.  
While deep learning techniques have been used for decades in HEP, their deployment for very low-latency (sub-microsecond) real-time particle tracking is just now being studied in depth, so there are many open questions about the best way to incorporate them. 
RNNs~\cite{Tsaris_2018} and Graph Neural Network (GNNs)~\cite{GNNs,Elabd:2021lgo,DeZoort:2021rbj,Ju_2021} have both shown great potential for identifying candidate tracks.  

In this paper we  explore the applicability of several advanced machine learning algorithms to different steps of tracking and compare their efficiency. One example of FPGA implementation is also shown.

\section{Machine Learning for Tracking}
\label{sec:MLfortracking}
The current generation tracking methods~\cite{Aaboud_2017} divides the problem into stages. 
These stages are: hit clustering, track seeding, track
extrapolation, track fitting and possibly removal of fake tracks. The hit clustering estimates the charge deposit and position parameters of particles that intercept with multiple pixels on a detector layer. Track seeding combines clusters in doublet or triplet seeds where each seed
provides an initial track direction, origin. The track
extrapolation stage adds more hits to the seed by
looking for matching hits along the extrapolated trajectory. Finally a track fitting stage  fits a trajectory through the track hits to assess the track
quality and measure the particle’s physical and kinematic properties (charge, momentum, origin, etc).
Current bottlenecks for offline tracking are the track seeding and extrapolation stage.
In this paper two different algorithms are studied for the track extrapolation part  and the removal of fake and duplicate tracks which are created at the track building stage.

\subsection{Extrapolator}
Traditional track extrapolation methods rely on a Kalman Filter (KF) to locate hits along the trajectory of charged particles in a detector \cite{KalmanFilter}. The Kalman Filter makes next-hit predictions via a state equation which propagates charged particles through potentially non-uniform magnetic fields and  detector materials. The predictions are combined with per hit measurements to progressively refine the predictions along a track. A series of matrix operations including addition, multiplication, and inversion are required by the Kalman filter at each iteration, making KF algorithms ill-suited for CPU driven computation. Some of the related problems may be addressed by heterogeneous computing strategies \cite{KalmanFilterGPU}. Additional constraints arise however, as magnetic field configurations and detector geometries must be stored and continually accessed in order for the KF to operate. The memory consumption of these steps further impedes the latency of KF tracking algorithms and strains memory limited devices such as FPGAs. 

Neural network based approaches to tracking provide natural solutions to these problems. 
Neural network architectures based on the multi-layer perceptron (MLP) rely on a series of matrix multiplications to perform tasks and so are amenable to firmware implementations on accelerating devices such as FPGA. Additionally, many feed forward neural networks have been shown to be universal approximators, and may be well suited to approximating functions such as the state equation for a charged particle propagating through a detector \cite{NNApproximators}. NNs also naturally address the issue of magnetic field and detector geometry storage and accessing. One of the many strengths of neural networks is their ability to extract features from a dataset and incorporate these features into the their predictions. Given sufficient training data of particle tracks, a NN can learn the magnetic field orientation and detector geometry of a generic particle detector. This eliminates the need to store and access such information independently on a chip.
 
Several models were developed for track extrapolation in this paper  and their performances are compared. Models are kept relatively narrow and shallow so that they may realistically fit with the resource budget of an FPGA. A description of the NNs models used is found below:

\subsubsection*{Multi-layer Perceptron (MLP)}
The simplest form of a neural network is the multi-layer perceptron (MLP). Given an input vector, of size $m$, $\mathbf{x}_1 = \mathbf{h}_1 \in \mathbb{R}^{m}$, we can compute the state layer vector at $l+1$ layer  $\mathbf{h}_{l+1} \in \mathbb{R}^{N_{l+1}}$, where $N_{l+1}$ is the dimension of the $l+1$ layer state space by iteratively applyg:  
\begin{equation}
\mathbf{h}_{l+1}(t)=\phi(\mathbf{W_{hl} \times h_l}(t)+\mathbf{b_{hl}})\label{eq:mlp}
\end{equation}
where $\mathbf{w_{hl}}$ is an $N_l \times N_{l+1}$ weight matrix and $\mathbf{b_{hl}}$ is the bias weight, and $\phi$ is a non-linear activation function such as ReLU, sigmoid, and tanh to generate nonlinear effects in the latent representation. The last layer of MLP is connected to the targeted output $\mathbf{y}(t)$ by additional activation or readout function.
The major advantage of the MLP is its simplicity. It is able to learn non-linear mappings, but it struggles to capture temporal or sequential dependencies of input data. The MLP used in this paper is comprised of an input layer, two hidden layers with ReLU activations, and an output layer. A tanh activation is applied to the final layer so that the outputs correspond accordingly with the scaled inputs.

\subsubsection*{Recurrent Neural Network (RNN)}

Recurrent neural network (RNN) is a type of neural networks that is designed to handle sequential or time series data. The basis of RNN is learning the \textit{recurrent} latent representation of the sequence or time series. Specifically, the temporal dynamics of hidden states are governed by a recurrent connection:  
\begin{equation}
\mathbf{h}(t)=\phi(\mathbf{W_{hh}h}(t-1)+\mathbf{W_{hx}x}(t)+\mathbf{b_h})\label{eq:rnn}
\end{equation}
where $\mathbf{W_{hh}}$ is an $N\times N$ weight matrix, $\mathbf{W_{hx}}$ is an $N\times K$ weight matrix, $\mathbf{b_{h}}$ is the bias weight, and $\phi$ is an activation function. The output $\mathbf{y}(t)\in \mathbb{R}^{L}$ is mapped from the hidden state: 
\begin{equation}
\mathbf{\hat{y}}(t)=f(\mathbf{h}(t))\label{eq:rnn2}
\end{equation}
where $f$ is a readout function related to the task usually set as a feedforward neural network. 

For the studies presented in this paper,  GRUs (Gated Recurrent Units) are chosen. These additionally contains a forget gate to cope with vanishing gradients but remains relatively simple, such that they may be effectively programmed onto an FPGA.

\subsubsection*{Bidirectional Recurrent Neural Network (biRNN)}
This model is identical to the RNN, however each input sequence is processed in both the forward and backwards direction. The output of each process is concatenated together before the output activation is applied. 

\subsubsection*{Echo-state Network (ESN)}

The computational cost of RNN can be expensive due to its inherently sequential nature and the overhead of avoiding the vanishing gradient problem. Echo state network (ESN) \cite{jaeger2007optimization} are designed to mitigate this. 

An ESN is built by connecting input vectors to a "reservoir" of nodes via a randomly initialized weight matrix. The output from this operation is multiplied by a sparse matrix which is also randomly initialized and held fixed during training. Finally, the reservoir is passed through a trainable feed forward neural network to give the desired output. 

The ESN used here have a reservoir of 200 nodes with a connectivity of 10\% and tanh activation.

\subsection{Overlap removal}
For any track reconstruction algorithm, even those that rely on Kalman Filter or Hough Transform~\cite{HoughTransformDuda,HoughTransformOrig} or the NN based algorithms described above, combinations of hits that do not match to any truth particles are commonly created. This arise due to the large combinatorial possibilities for initial seeds from which tracks are extrapolated from or when a possible hits are searched for in the next layers. The number of `fake' tracks typically scale with the density of particle in an event, and can reach as high as approximately 1000 fake tracks created for each `true' track at conditions expected at the HL-LHC \cite{ATLASTDR}.

These fake tracks are typically rejected by performing a full track fit and performing a selection on the compatibility of the hits to originate from a potential particle traversing the detector. However, these fits are computationally expensive and require a precise description of the magnetic field and the detector geometry. 

Similar to the requirements of the track extrapolator, NNs offer a simple way to condense the information such that they can be implemented on memory limited devices. Additionally, the expected time required to evaluate an NN on a collection of hits is significantly less than that of a full track fit. This allows a significant fraction of fake hit combination to be rejected before a full track fit is performed.

\section{Samples}
\label{sec:samples}
The simulated samples and tracking data used in study are produced using the A Common Tracking Software Toolkit (ACTS) \cite{ACTS} using its fast track simulation engine. A generic detector geometry, OpenDataDetector (ODD) \cite{corentin_allaire_2022_6445359}, is used (shown in Figure~\ref{fig:predictedHitsOverlay}. It consists of the typical silicon technologies that are expected to be used at HL-LHC detectors. The detector has barrel and end-cap sections, with 4 (7) layers of pixel sensor placed closest to the interaction point in the barrel (end-cap) region. These are subsequently followed with 4 layers of short-strip silicon sensors and 2 layers of double-sided long-string silicon sensors in both regions. 

Various processes are simulated using this setup. For training the various ML algorithms, predominately events with single muons with a flat \pt\ distribution above 1~\GeV\ and no restriction on $\eta,\phi$ of the track are used. These samples are supplemented with simulated proton-proton collision events, where the hard-scatter process is forced to only create a top anti-top pair. The decay of these particles provides a large variety of charged particles across different \pt\ regions. In addition to being used in the training dataset, this sample is used to study the number of hits, clusters and roads that the system will encounter in data-taking.

The single muon and top anti-top events are also overlaid with soft QCD ``pileup'' vertices to simulate the highly dense environment observed in proton--proton collision. The average number of vertices, $<\mu>$, are set to 40 and 200 to simulate the current LHC condition and the future expected HL-LHC condition respectively.


\section{Track reconstruction}
\label{sec:trackReco}

The first step of track reconstruction is the formation of hit clusters from raw hit data, the goal being to group adjacent hits and average their positions to result in one coordinate per detector layer.
These clusters are then used to perform track reconstruction,
via a first high-speed but low precision step of track seeding.
In this paper, we compare the performance of a track seeding 
algorithm under investigation by the ATLAS~\cite{ATLASTDR},
known as the Hough Transform~\cite{HoughTransformDuda,HoughTransformOrig},
with track reconstruction performed by neural networks.

\subsection{Hough Transform}
The Hough Transform implemented here was inspired by work done for the EF Tracking Amendment of the ATLAS Trigger and DAQ upgdade Technical Design Report~\cite{ATLASTDR}.
Most concepts implemented there -- detector layers used, rough size of accumulator and bin sizes in $\phi_0$ and $qA/p_\mathrm{T}$, use of hit extension in inner layers for larger $d_0$ acceptance, $z$-slicing for reducing accumulator occupancy -- are also adopted here.
However, it was prepared for the under-construction ATLAS ITK, which has a different detector geometry to the ODD detector,
so several aspects of it are modified.
Full details can be found in Appendix~\ref{app:houghtransformdetails}. 

As identified in Ref~\cite{ATLASTDR} and also with the 
implementation used for the results in this paper,  
in order to obtain a high efficiency 
for track reconstruction with the Hough Transform, 
the selection requirements must be sufficiently loose. This generates a large number of fake tracks.
One avenue to improve the purity of the algorithm, are the studies presented in Section~\ref{sec:overlap}. A second way to improve this is to perform the track 
reconstruction itself with NNs, as described in section~\ref{sec:trackseeding}.

\subsection{Track Reconstruction with NNs}
\label{sec:trackseeding}
A set of NN's of increasing complexity are trained with the goal of finding the next hit position along a track, given initial hits along the track. The models used for this study are presented in Table \ref{table:modelCompare}. Each NN is given three sequential hits as input. The input features for each hit are the hit's x/y/z coordinate, with each element scaled to be O(1), as well as a one-hot-encoded vector specifying the detector volume and layer where the hit is located. The NN's are pre-trained with mean-square-error loss function  with the corresponding next hit's cartesian coordinates as a target,
\begin{equation}
MSE = \frac{1}{n}\sum_{i=1}^{n} (r_{i} - \hat{r}_{i})^{2}
\label{eqn:MSE}
\end{equation}

\begin{equation}
\mathcal{L} = MSE + \frac{A}{n}\sum_{i=1}^{n    } (\rho_{i} - \hat{\rho}_{i})^{2}
\label{eqn:mMSE}
\end{equation}

All models are trained over a set of nearly 3 million tracks, segmented into three hit sections used as input. Each track is comprised of only a single hit per layer. In cases when there are multiple hits in a single layer, the hit closest to the origin is used to construct a clean track. A batch size of 4096 is used for each training and the models are trained until the loss values converge over a set of validation samples. A RelU activation as applied to each hidden layer. Outputs are passed to a tanh function to ensure the outputs are O(1) in accordance with then normalization scheme.

\begin{figure}
    \centering
    \includegraphics[width=0.49\textwidth]{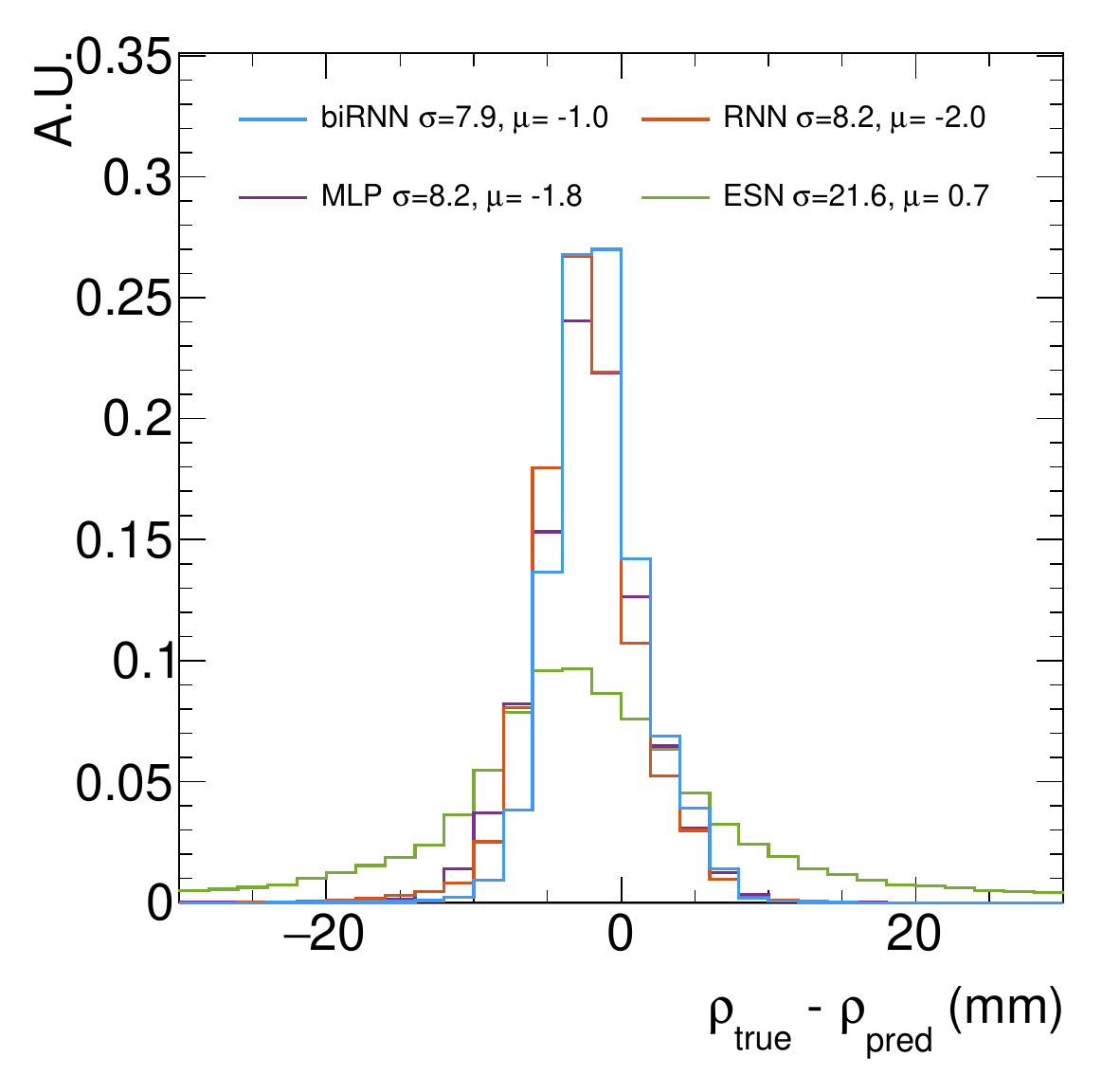}
    \includegraphics[width=0.49\textwidth]{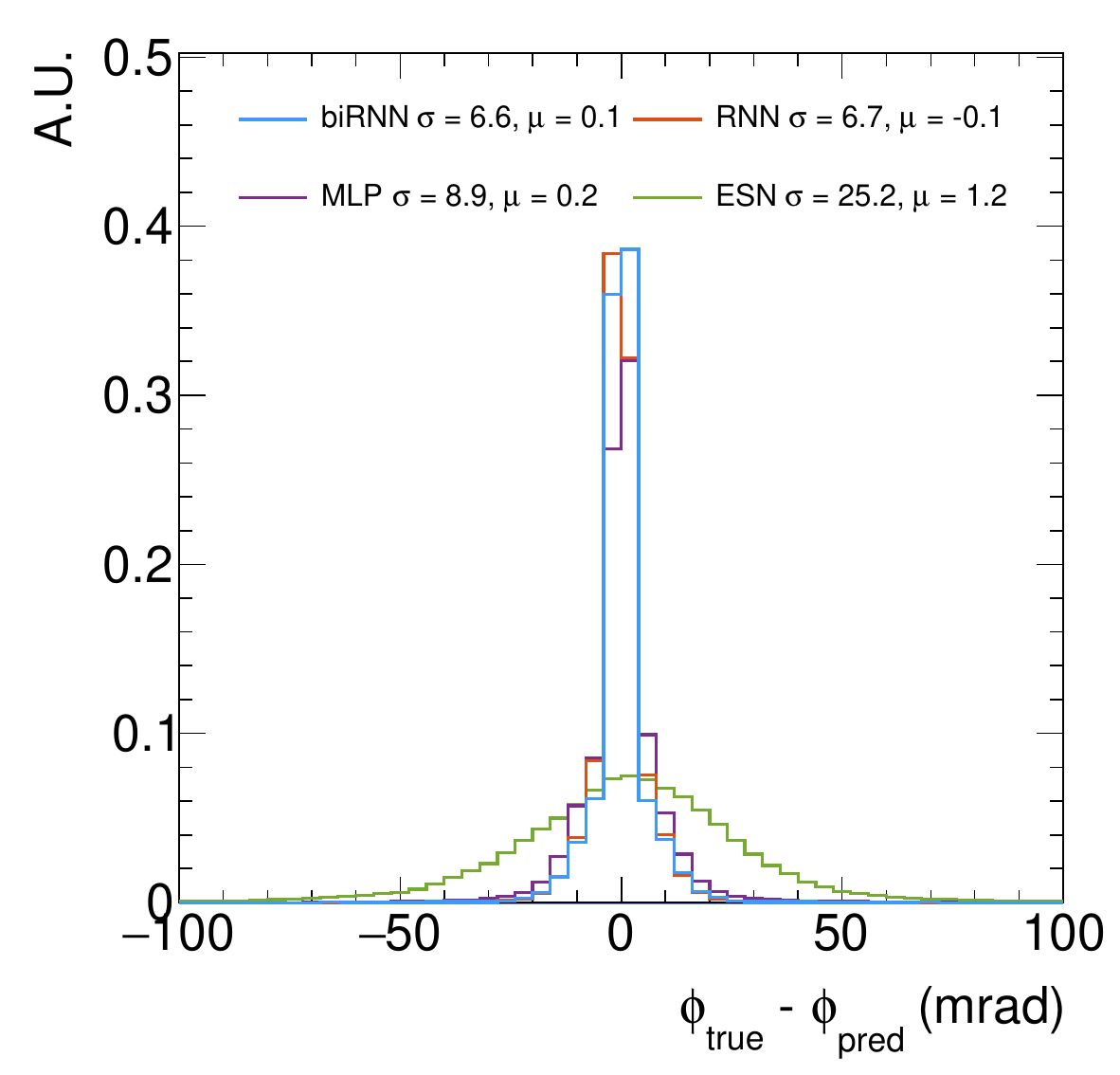}
    \includegraphics[width=0.49\textwidth]{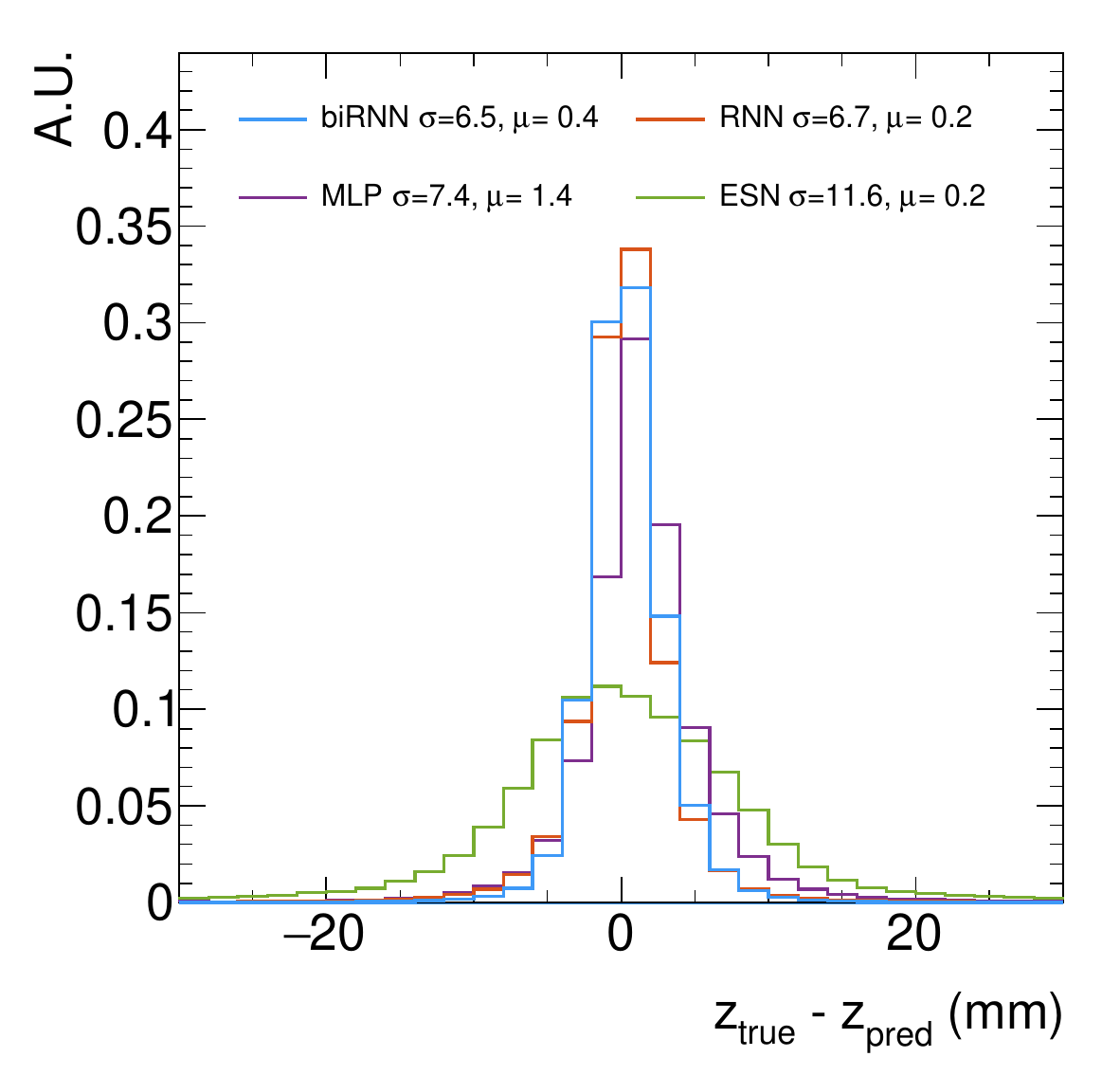}
    \caption{Overlaid residual distributions in cylindrical coordinates from the four neural networks. The standard deviation $\sigma$ and mean $\mu$ of each distribution are cited in the legend}
    \label{fig:residuals}
\end{figure}

\begin{figure}
    \centering
    \includegraphics[width=0.49\textwidth]{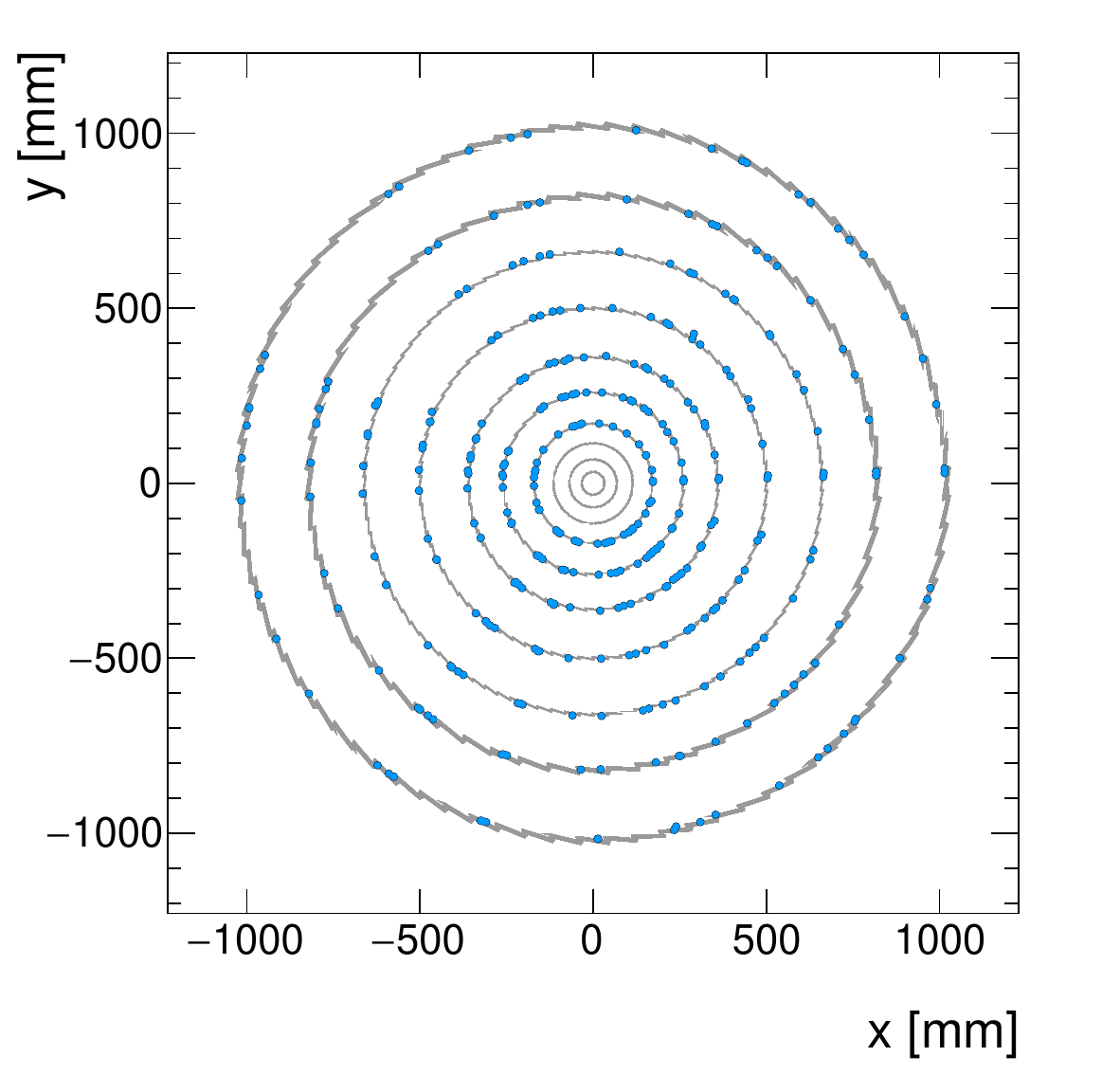}
    \includegraphics[width=0.49\textwidth]{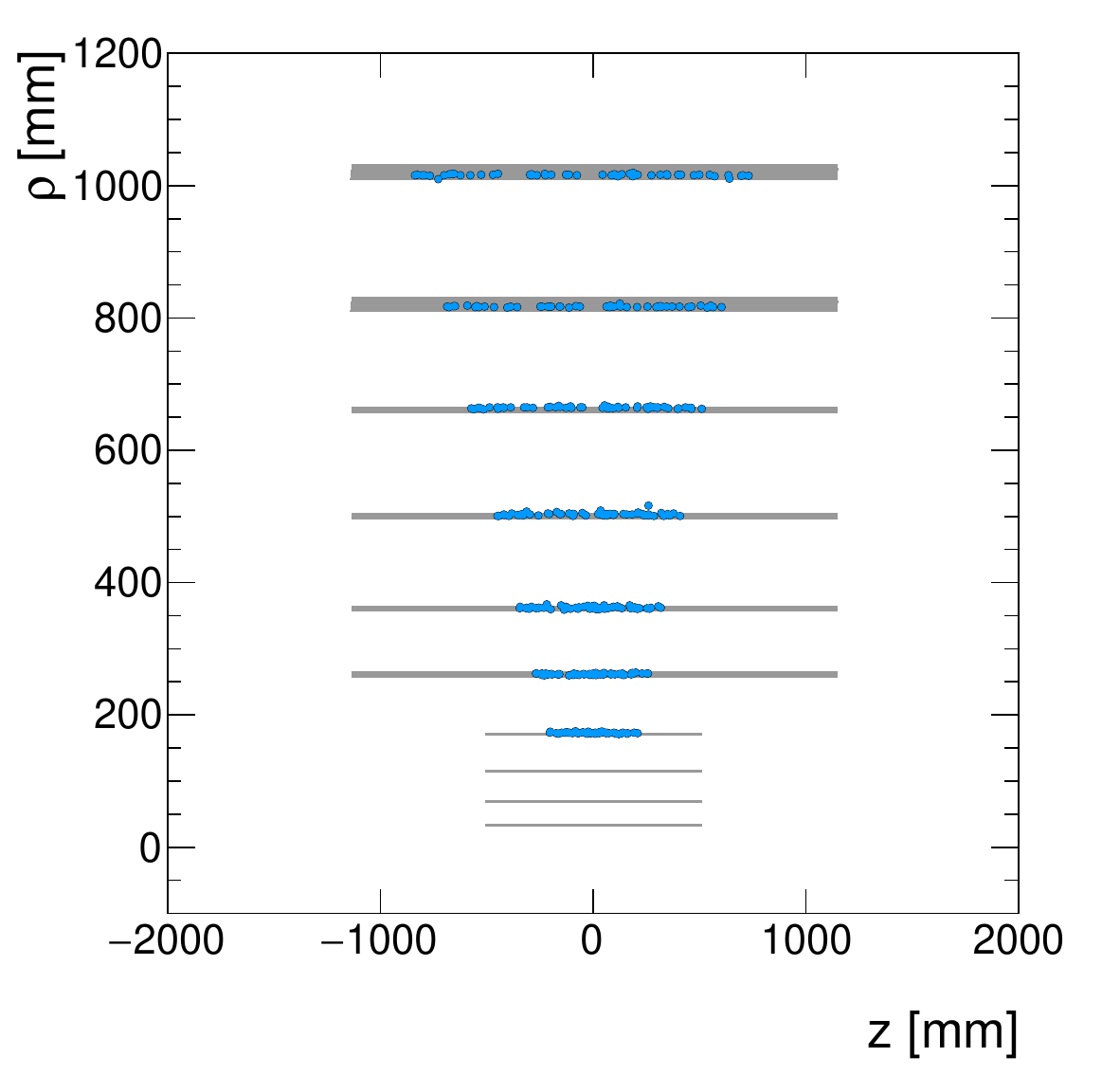}
    \caption{The predicted hit coordinates output by an RNN where the preceding three true hits are used as input, are overlaid on the ACTS detector. Tracks are required to have at least 8 hits within the barrel. \label{fig:predictedHitsOverlay}}
\end{figure}

The performance of the various NN is evaluated by comparing the residual between the true hit coordinates and the predicted coordinated shown in Figure~\ref{fig:residuals} and summarized in Table~\ref{table:modelCompare}. For a similar architectural complexity, the MLP, RNN and biRNN NN have better performance when compared to the ESN NN. Additional, to cross-check that NN have learned the detector geometry during the training, the NN predicted hits using the RNN model are overlaid on the various detector geometry as show in Figure ~\ref{fig:predictedHitsOverlay}. For most predicted hits, the coordinates lie within a detector element, with only a small fraction of hits lying outside the elements in the $\rho$ coordinate.

\begin{table}[h!]
\centering
\caption{Comparison of various neural network model performance attributes. The model architecture is read as the number of nodes in a given successive layer. \label{table:modelCompare}}
\begin{tabular}{ l|c|c|c|c|c  }
 \toprule
\multirow{2}{*}{NN Type} & \multirow{2}{*}{Model Architecture} & Training time & \multicolumn{3}{c} {$\sigma$ of residual}  \\
                         &                                     & per epoch [s]     & $\rho$ [mm] & $\phi$ [mrad]& z [mm]\\
 \midrule
 MLP   & 14 x 32 x 32 x 32 x 3 & 5   & 8.2 & 9  & 7.4\\
 RNN   & 14 x 32 x 32 x 3      & 9.5 & 8.2 & 7  & 6.7 \\
 biRNN & 14 x 32 x 32 x 3      & 15  & 7.9 & 7  & 6.5 \\
 ESN   & 14 x 200 x 3          & 6   & 21  & 25 & 11 \\
 \bottomrule
\end{tabular}
\end{table}

\subsubsection{Extrapolation Algorithm}
The extrapolation algorithm starts by operating on seed tracks, which are comprised of three hits in the inner-most layers. The NN uses these hits to predict the coordinate of the next hit along each track. A new track is constructed for each hit that lies within a given radius of the predicted coordinate. This radius is informed by the uncertainty of each NNs predictions at any given detector region. Three constant values of 10mm, 15mm, and 20mm are used throughout this paper as points of comparison. 

The most recent three hits along each extrapolated track are iteratively used as input to the NN, and tracks are constructed this way until either:
\begin{itemize}
    \item there are no hits in the detector within range of the predicted hit, or
    \item the edge of the detector is reached.
\end{itemize}

Many of the duplicate and fake tracks are created through this process are rejected via the method described in Section \ref{sec:overlap}.

\subsubsection{Performance}
\begin{figure}
    \centering
    \includegraphics[width=0.49\textwidth]{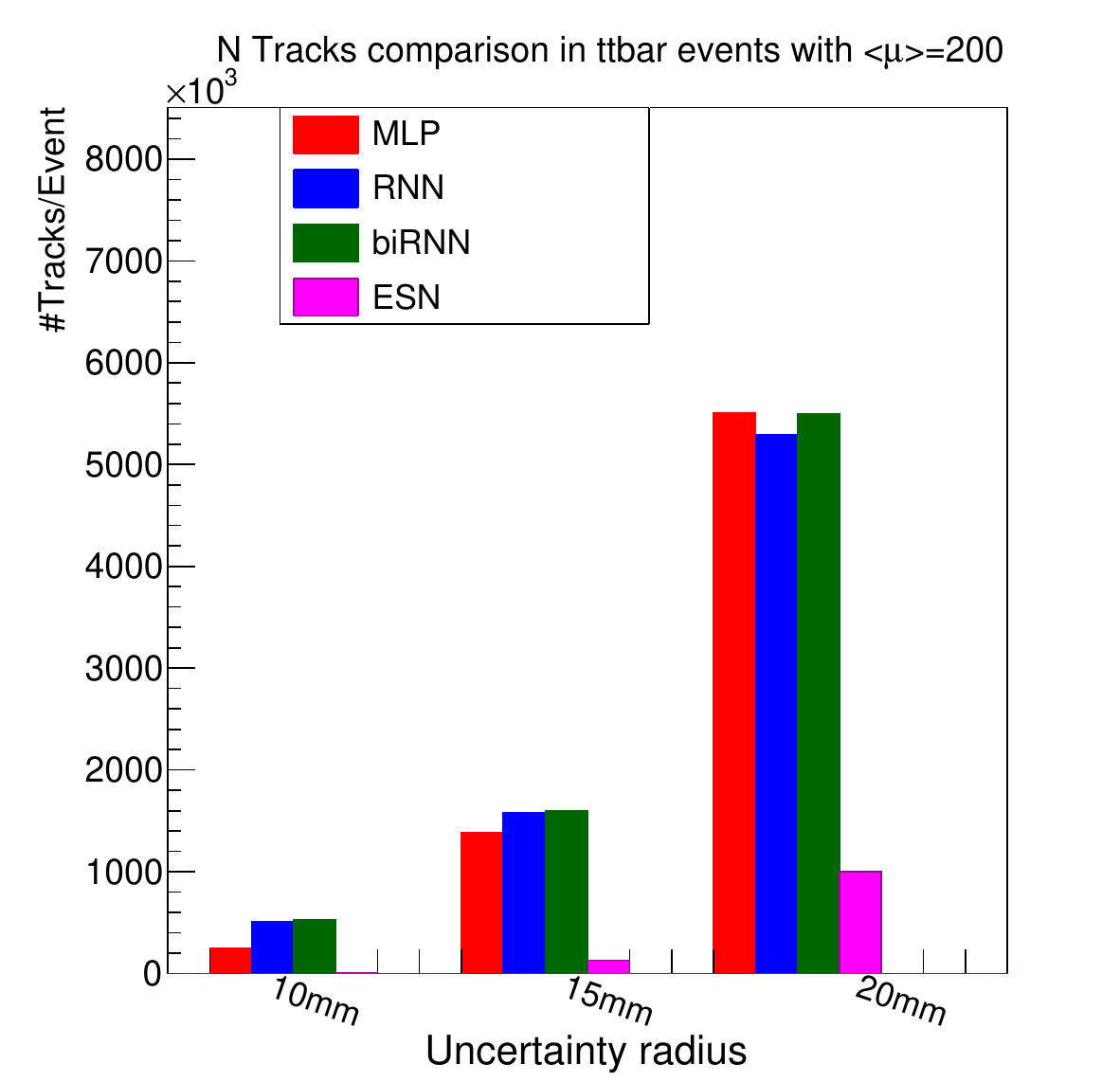}
    \includegraphics[width=0.49\textwidth]{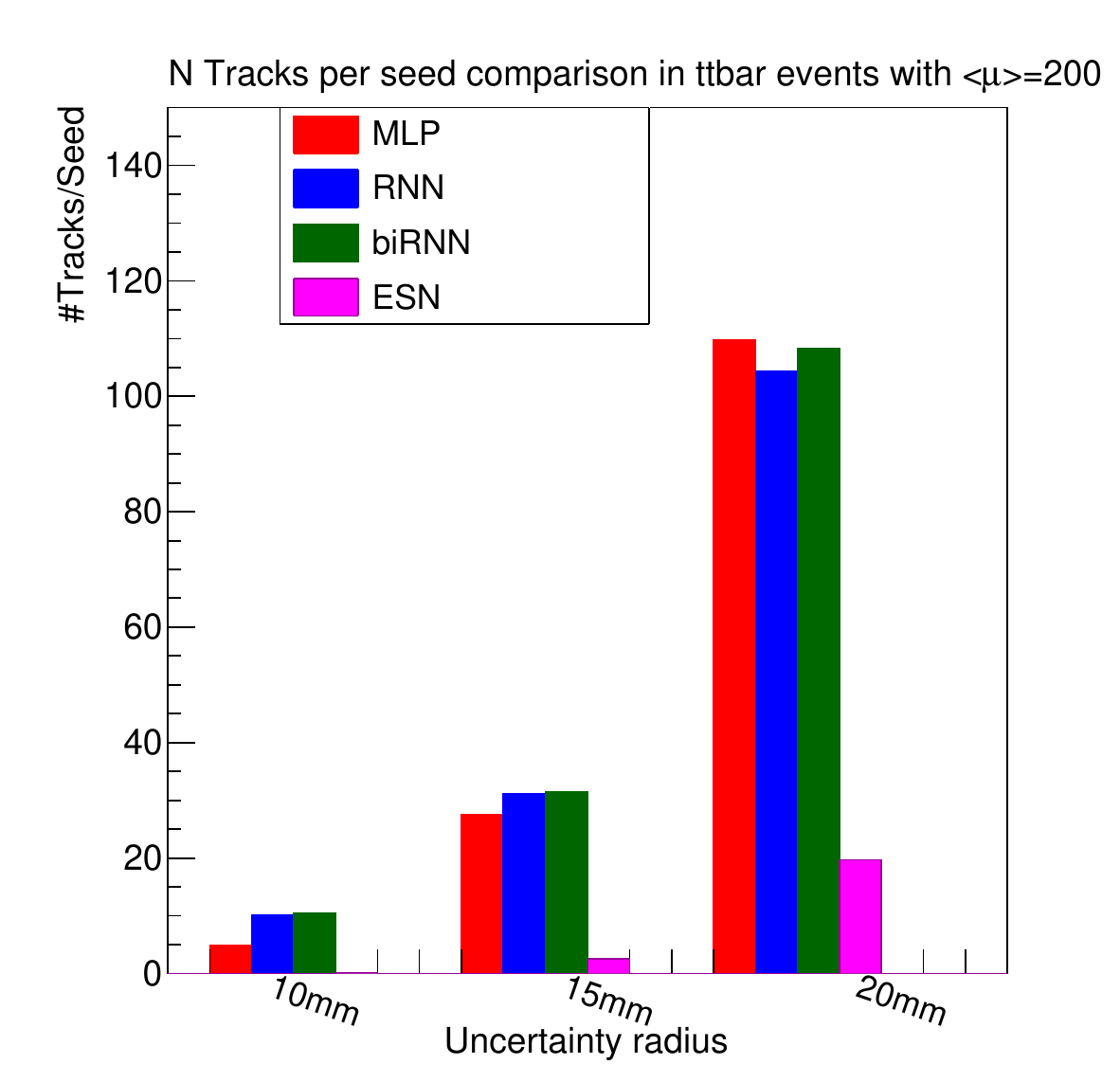}    \caption{The average number of total extrapolated tracks and the average number of extrapolated tracks for each seed track in a event with $<\mu> = 200 $ pileup event for various uncertainty window size.  \label{fig:comparison}}
    
\end{figure}

\begin{figure}
    \centering
    \includegraphics[width=0.49\textwidth]{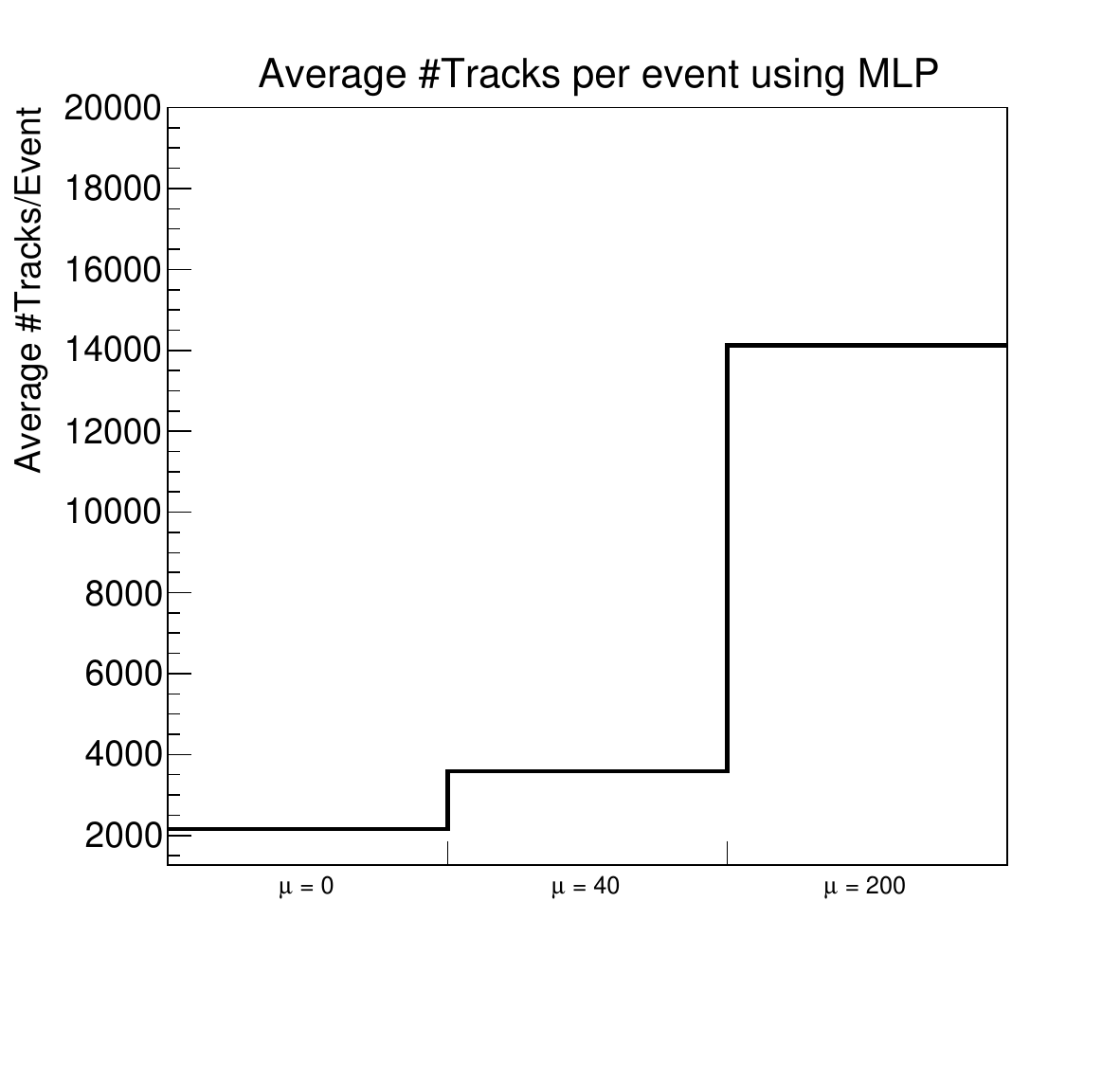}
    \caption{The average number of total extrapolated tracks for various pileup scenarios for a window size of 15mm.  \label{fig:PileUpComparison}}
\end{figure}

To evaluate the performance of the extrapolation algorithm, seed tracks are constructed from the truth tracks. This simplification leads to the most straight forward method for evaluating the performance of the algorithm as it will not artificially decrease the purity of the reconstructed tracks by including the fake tracks from fake seeds. The average number of total tracks reconstructed in an event and the average number of tracks reconstructed per seed track for different uncertainty window is presented in Figure \ref{fig:comparison}. The large number of combination are due to high density of hits in the detector from lower energy particles, and increases with the average pileup in the event as showing in Figure \ref{fig:PileUpComparison}.

\begin{figure}
    \centering
    \includegraphics[width=0.49\textwidth]{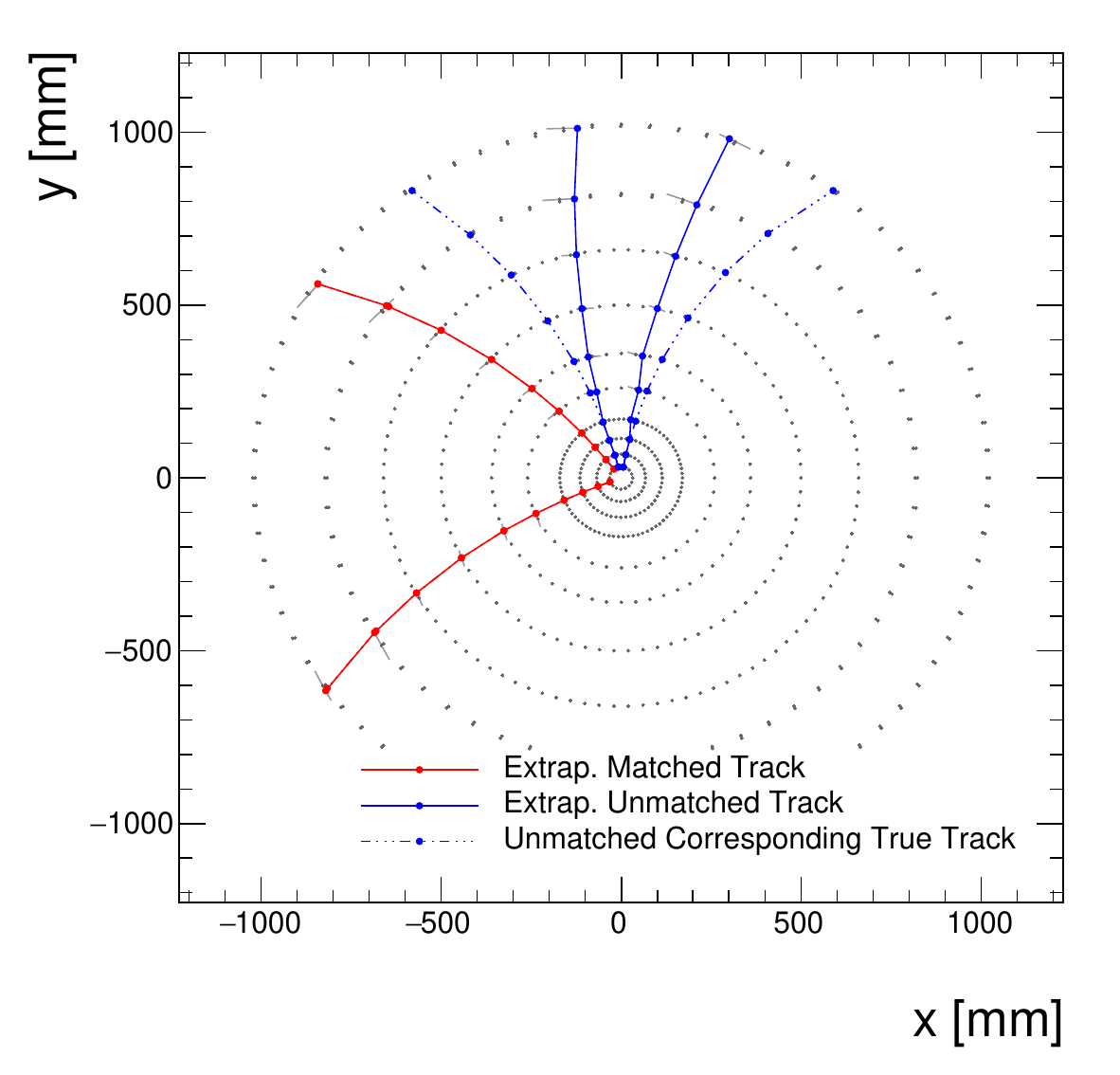}
    \includegraphics[width=0.49\textwidth]{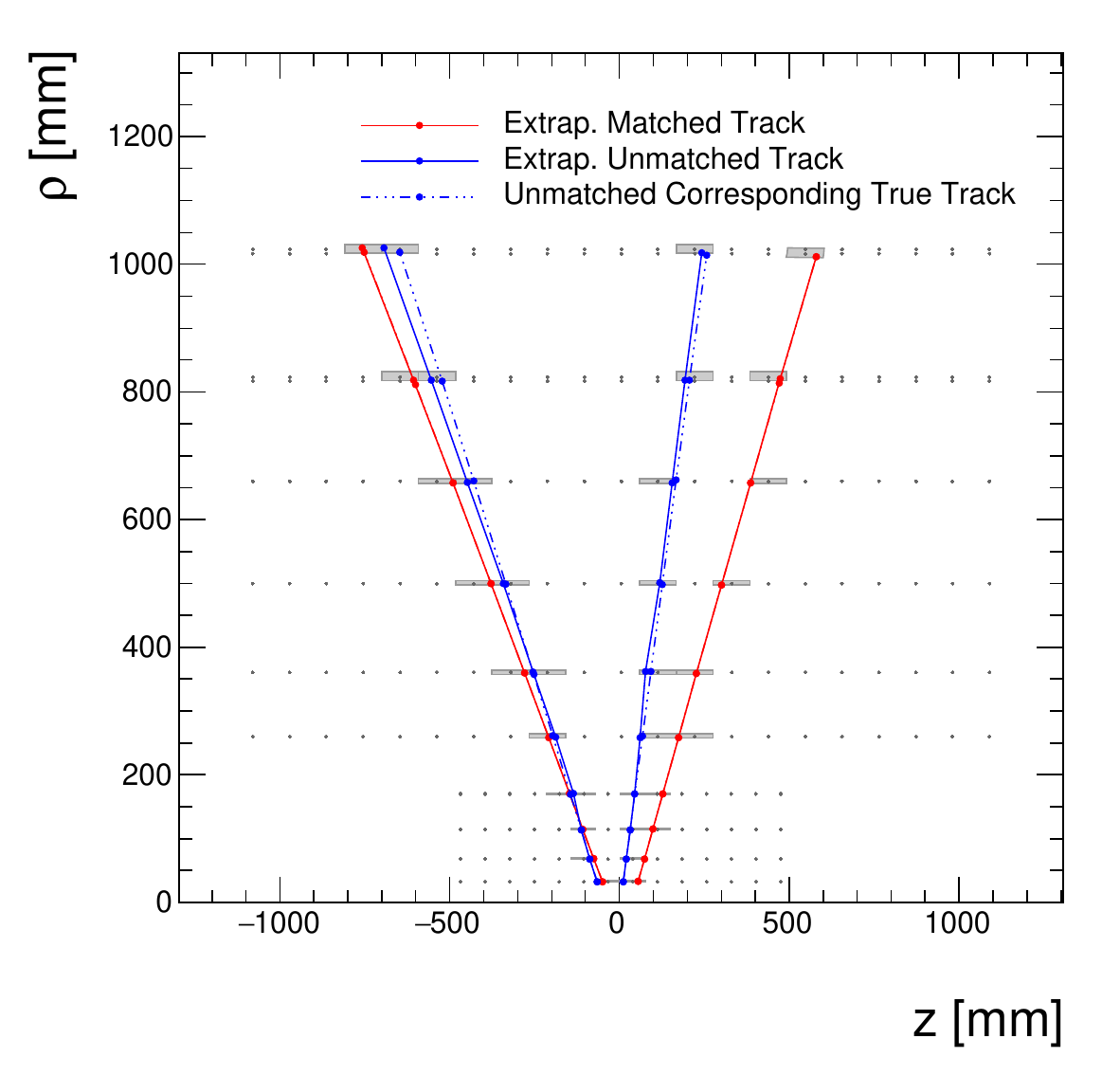}
    \caption{Four sample seeds and corresponding extrapolated tracks in a $t\Bar{t}$ event. Fully extrapolated tracks that are matched to a true track are shown as solid red lines. Extrapolated tracks that are not matched to a true track are shown as solid blue lines. The true track corresponding to the seed of the unmatched extrapolated tracks are shown as dashed blue lines. }
    \label{fig:tracksdetector}
\end{figure}

\begin{figure}
    \centering
    \includegraphics[width=0.45\textwidth]{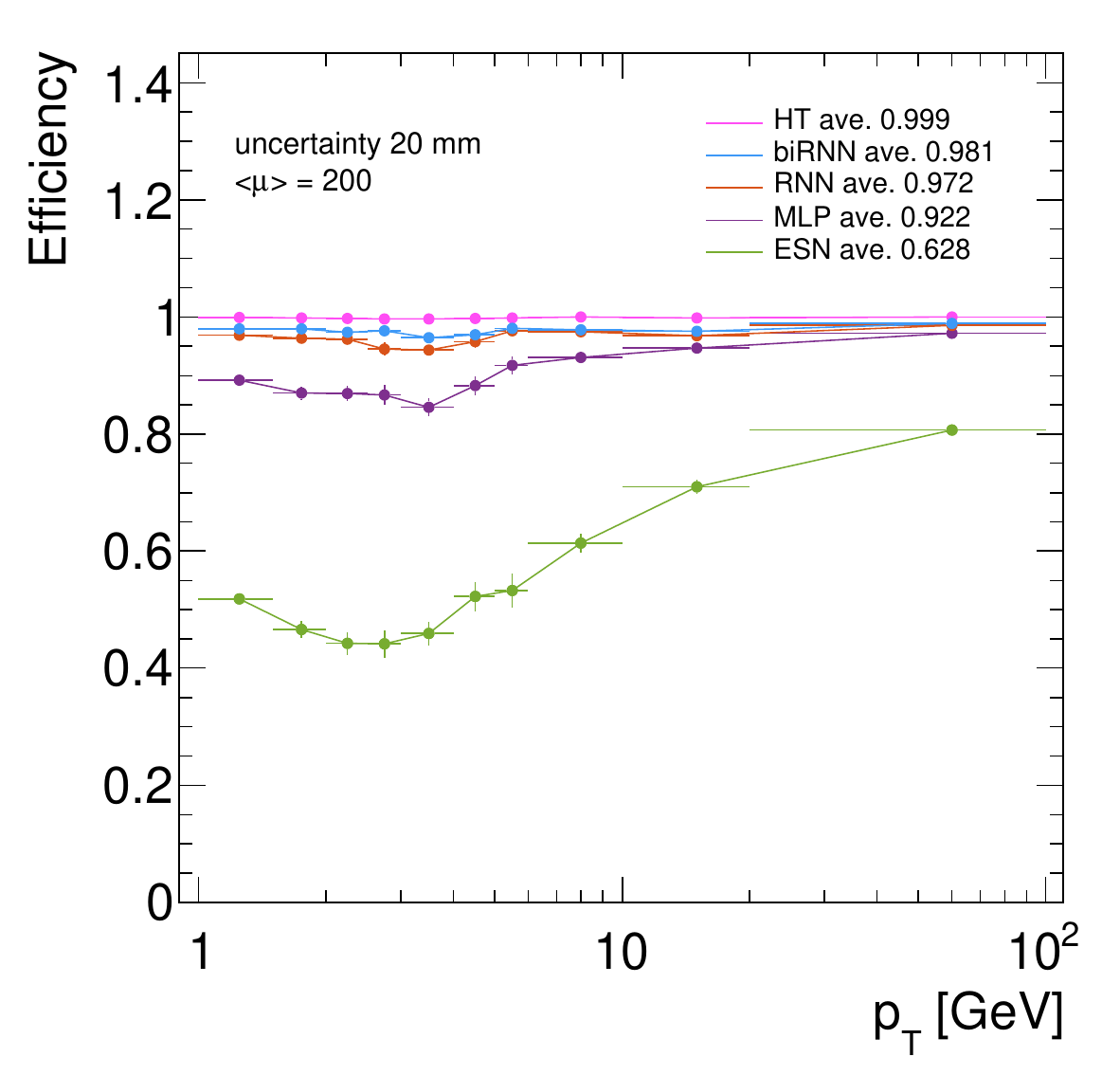}
    \includegraphics[width=0.45\textwidth]{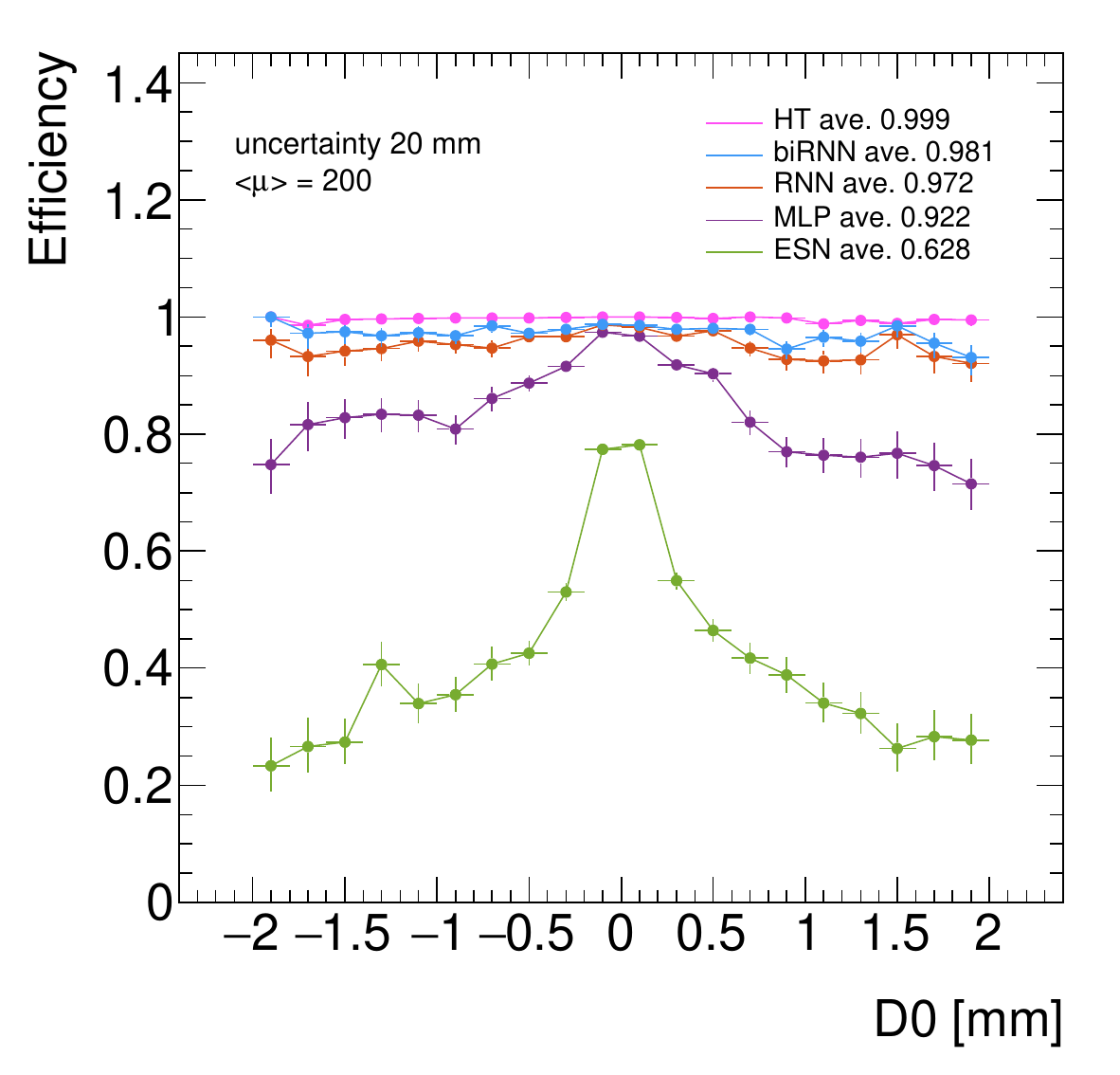}
     \includegraphics[width=0.45\textwidth]{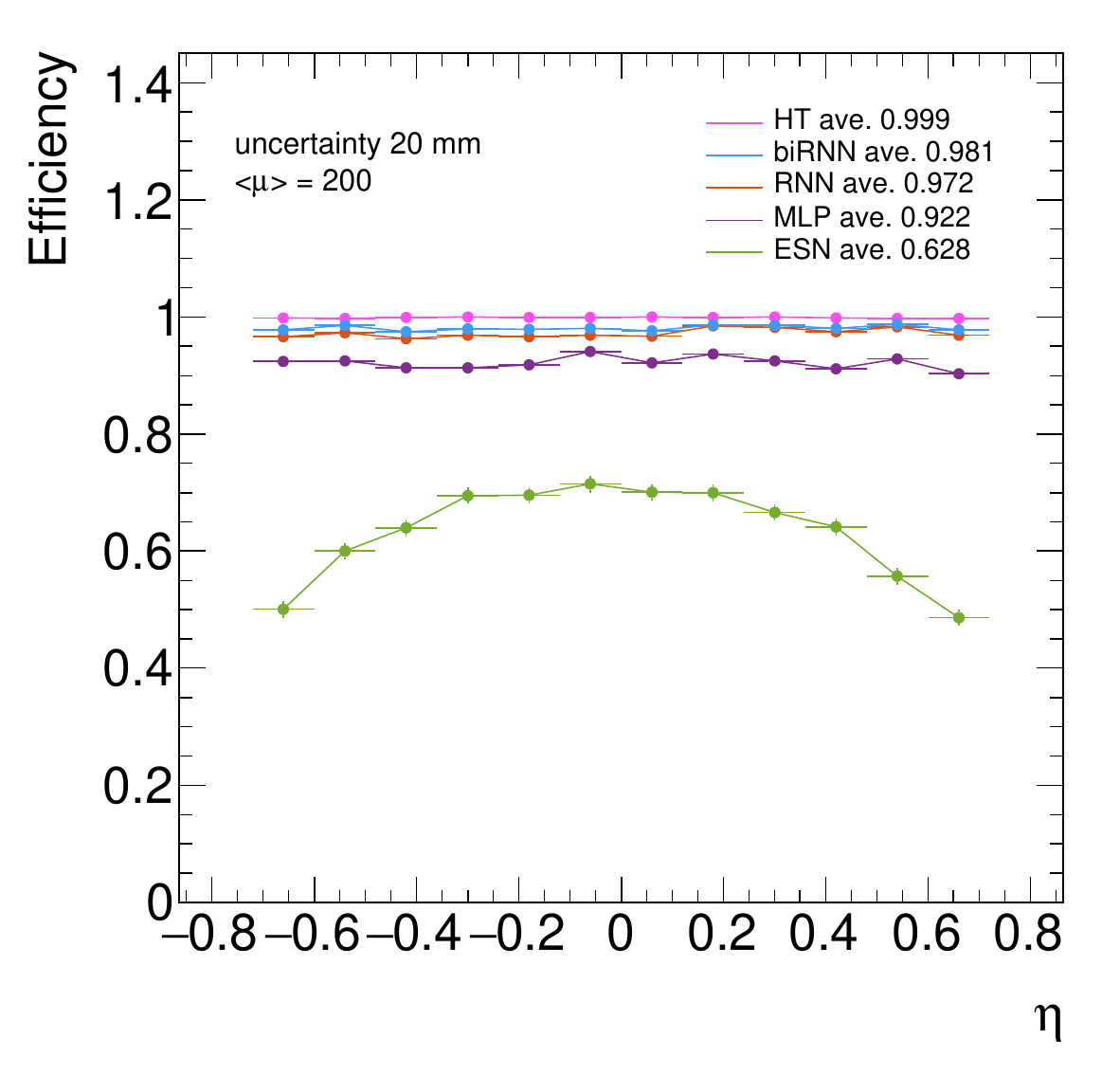}
     \includegraphics[width=0.45\textwidth]{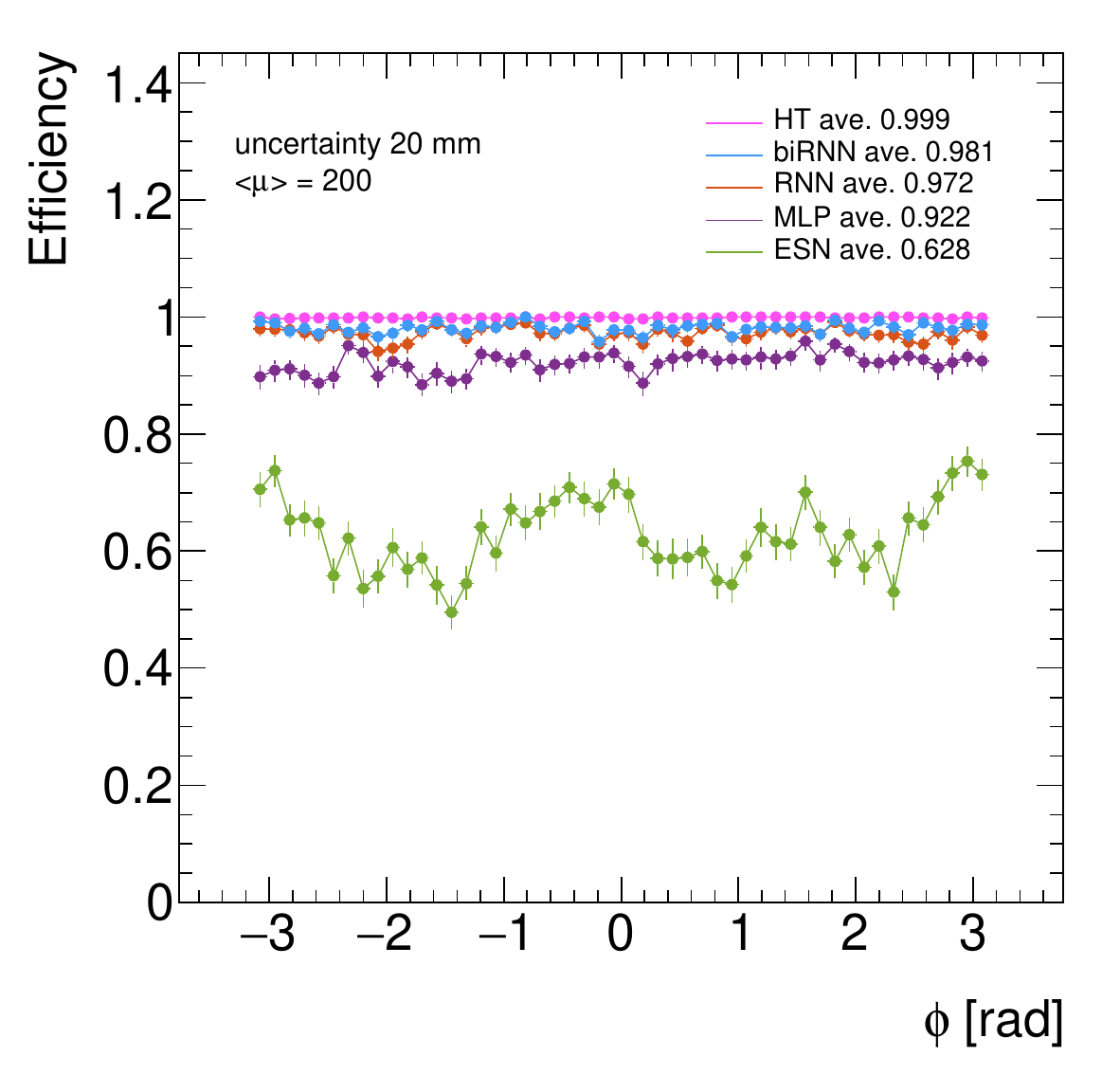}
     \includegraphics[width=0.45\textwidth]{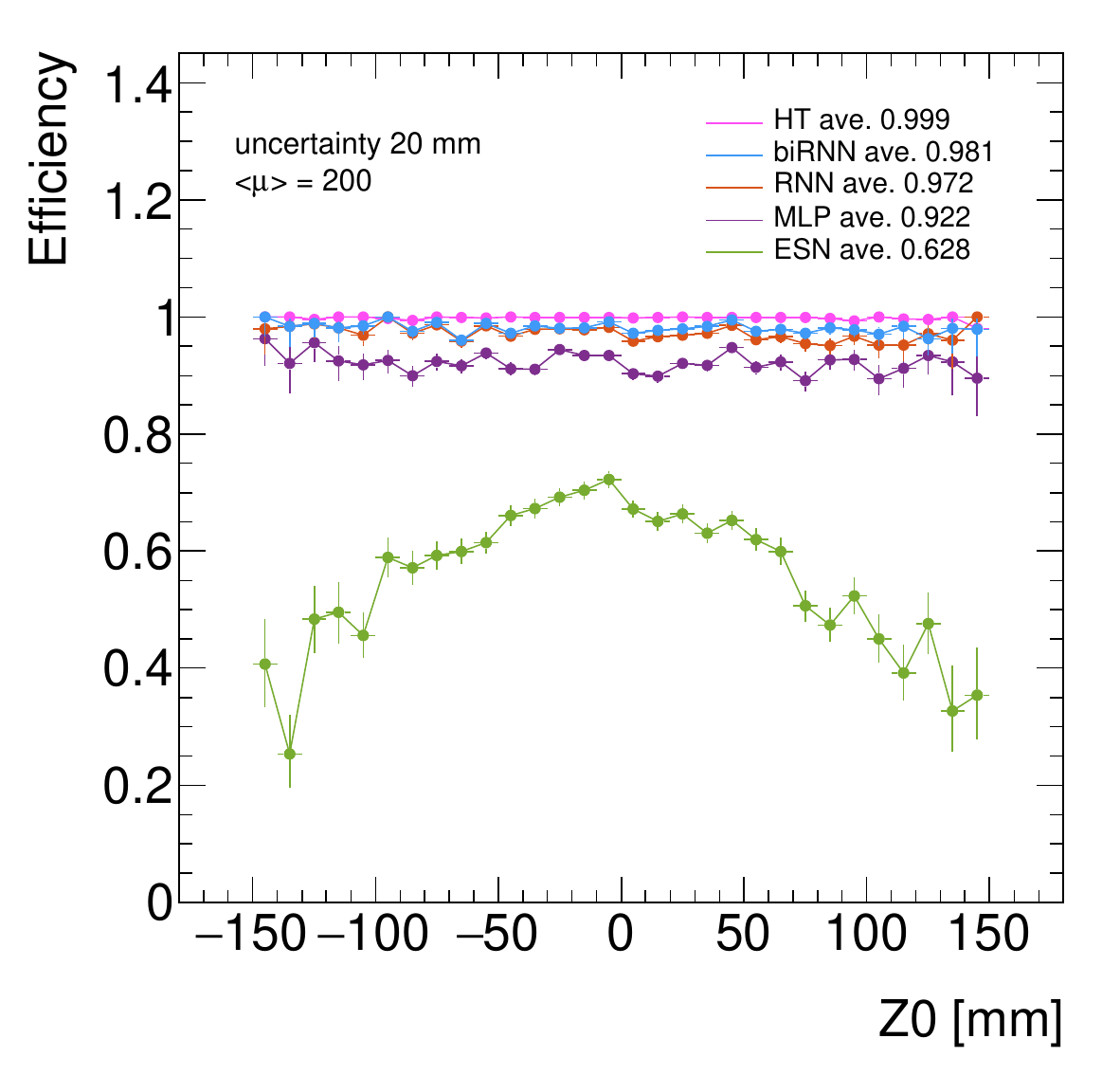}

    \caption{Reconstruction efficiency for different NN algorithms. Only tracks with $|\eta|<0.7$ are considered. The true and extrapolated tracks are required to have a minimum of 8 hits. An extrapolated track is considered "matched" to a true track if more than 50\% of the hits in the extrapolated track, excluding seed hits, stem from the same particle. The reconstruction efficiency is calculated as the fraction of truth tracks that are matched to at least one extrapolated track
    within an uncertainty window of 20mm.}
    \label{fig:efficiencyoverlay20}
\end{figure}

\begin{figure}
    \centering
    \includegraphics[width=0.45\textwidth]{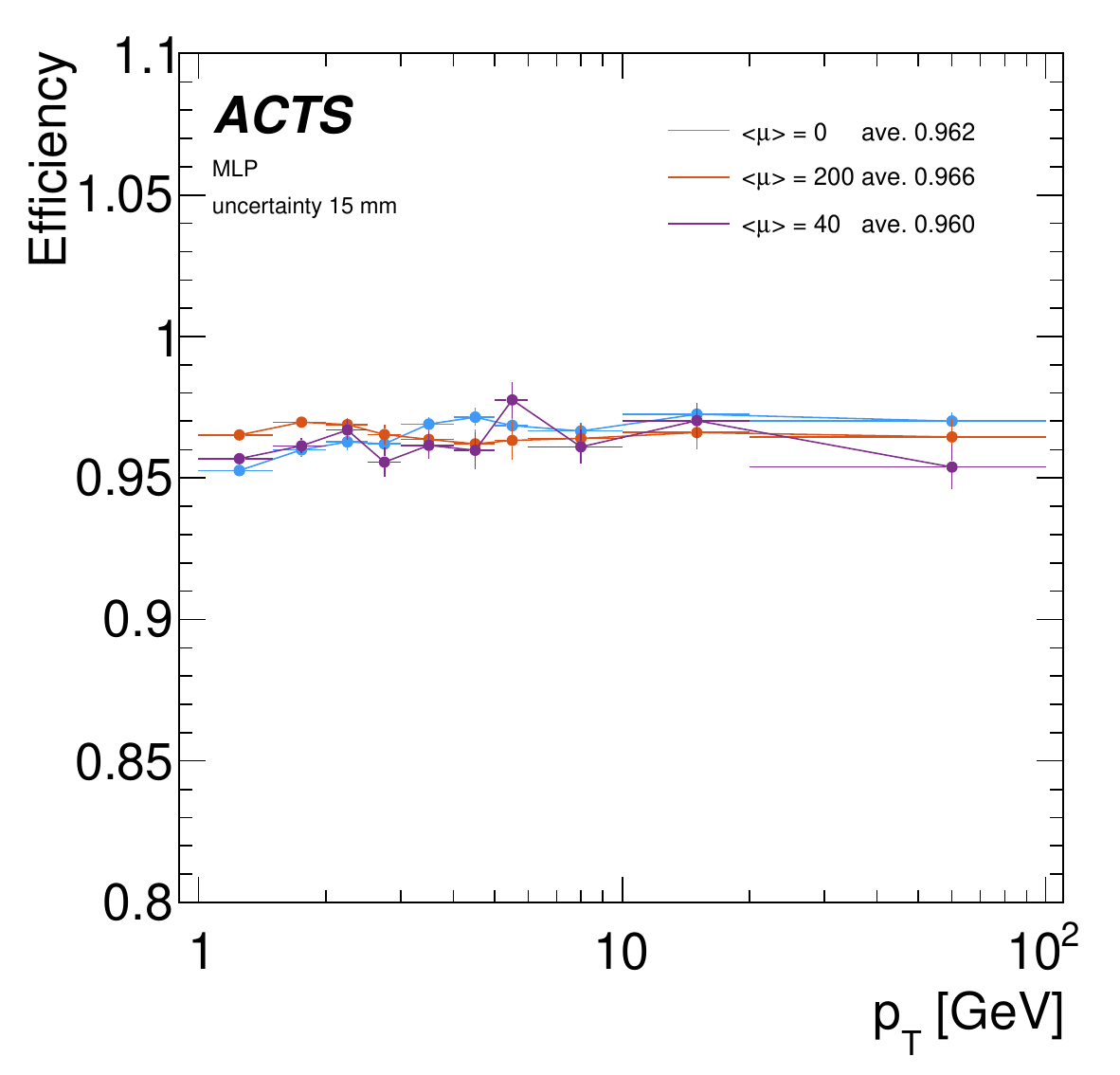}
    \includegraphics[width=0.45\textwidth]{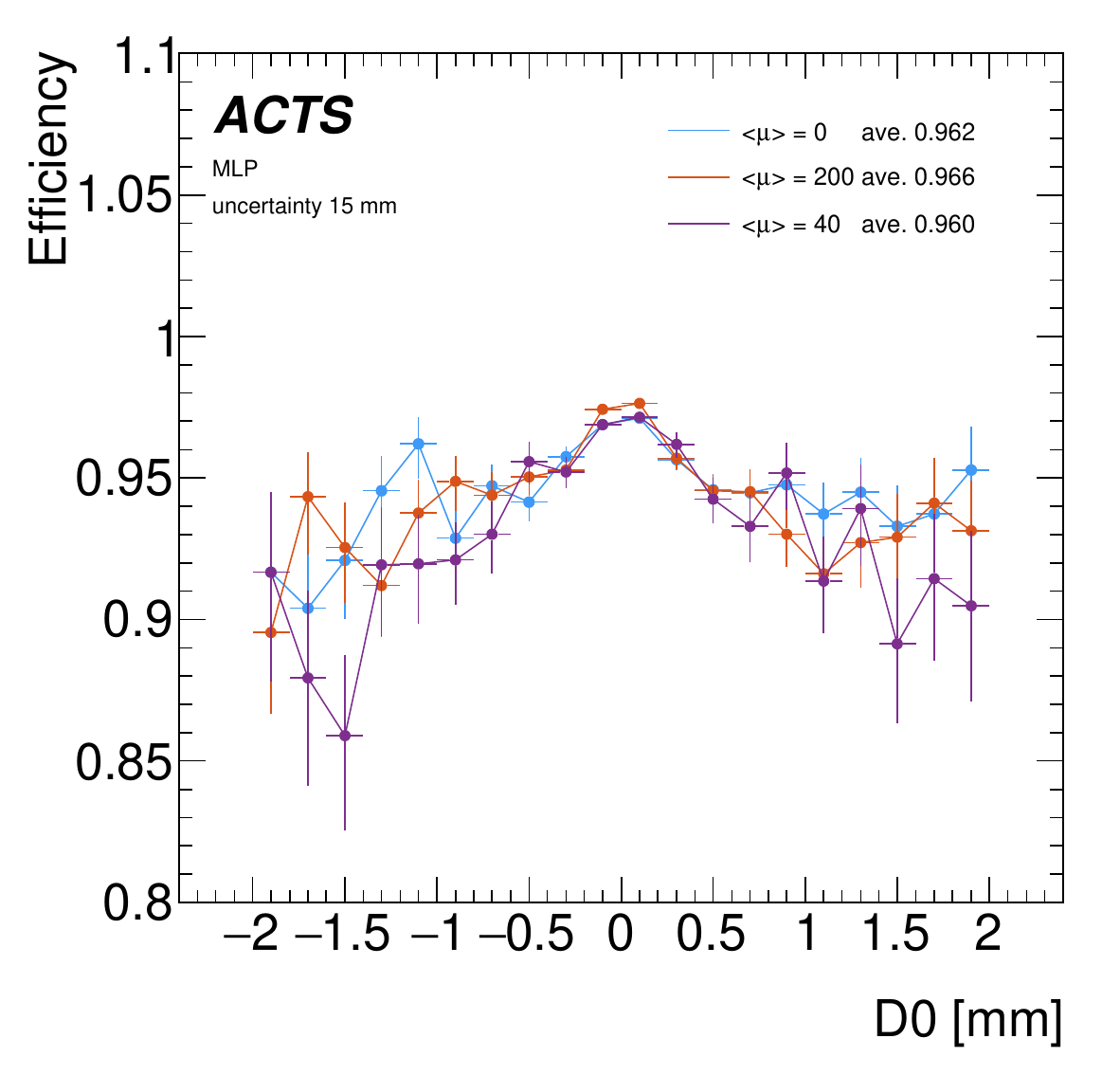}
     \includegraphics[width=0.45\textwidth]{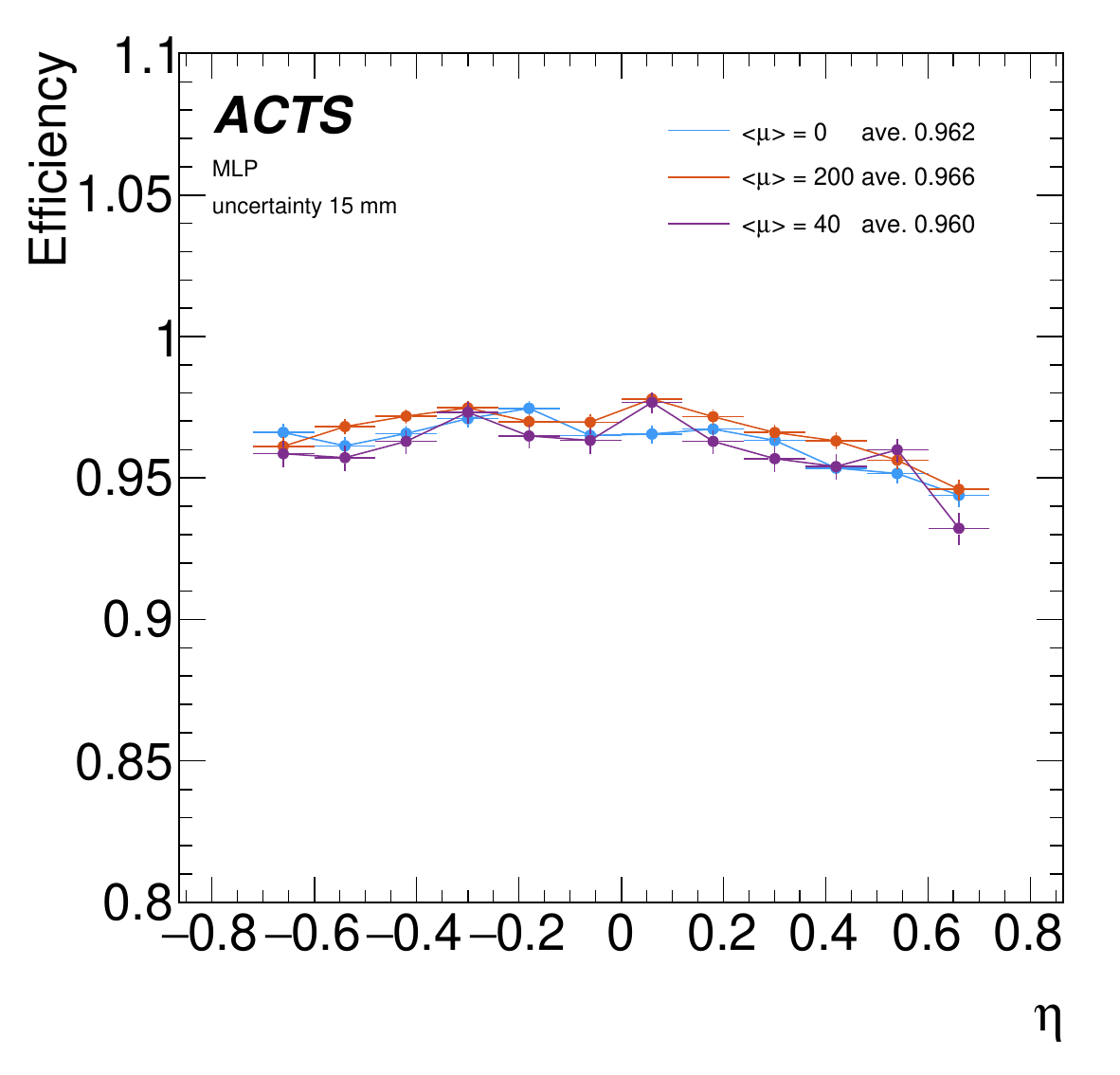}
     \includegraphics[width=0.45\textwidth]{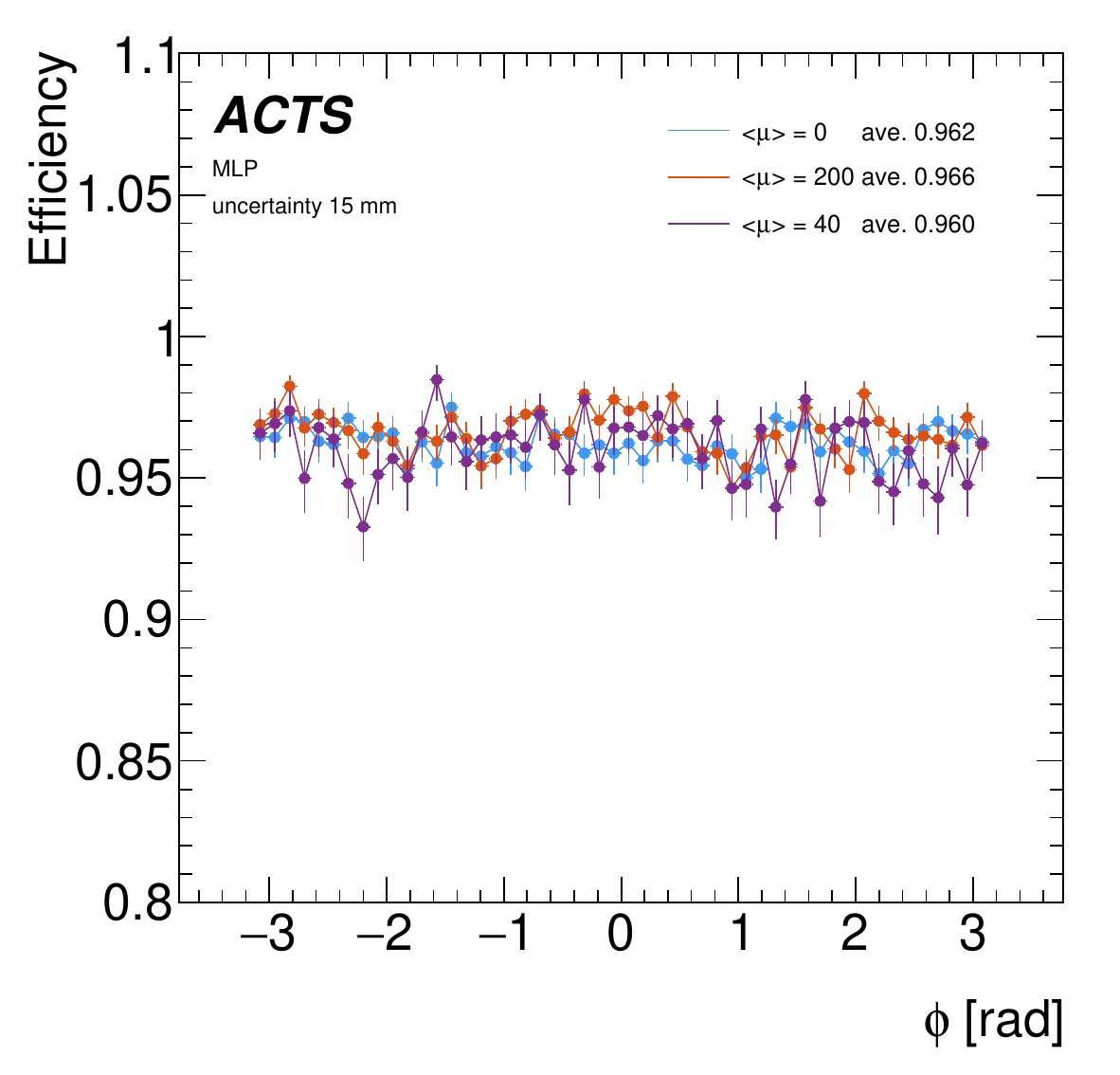}
     \includegraphics[width=0.45\textwidth]{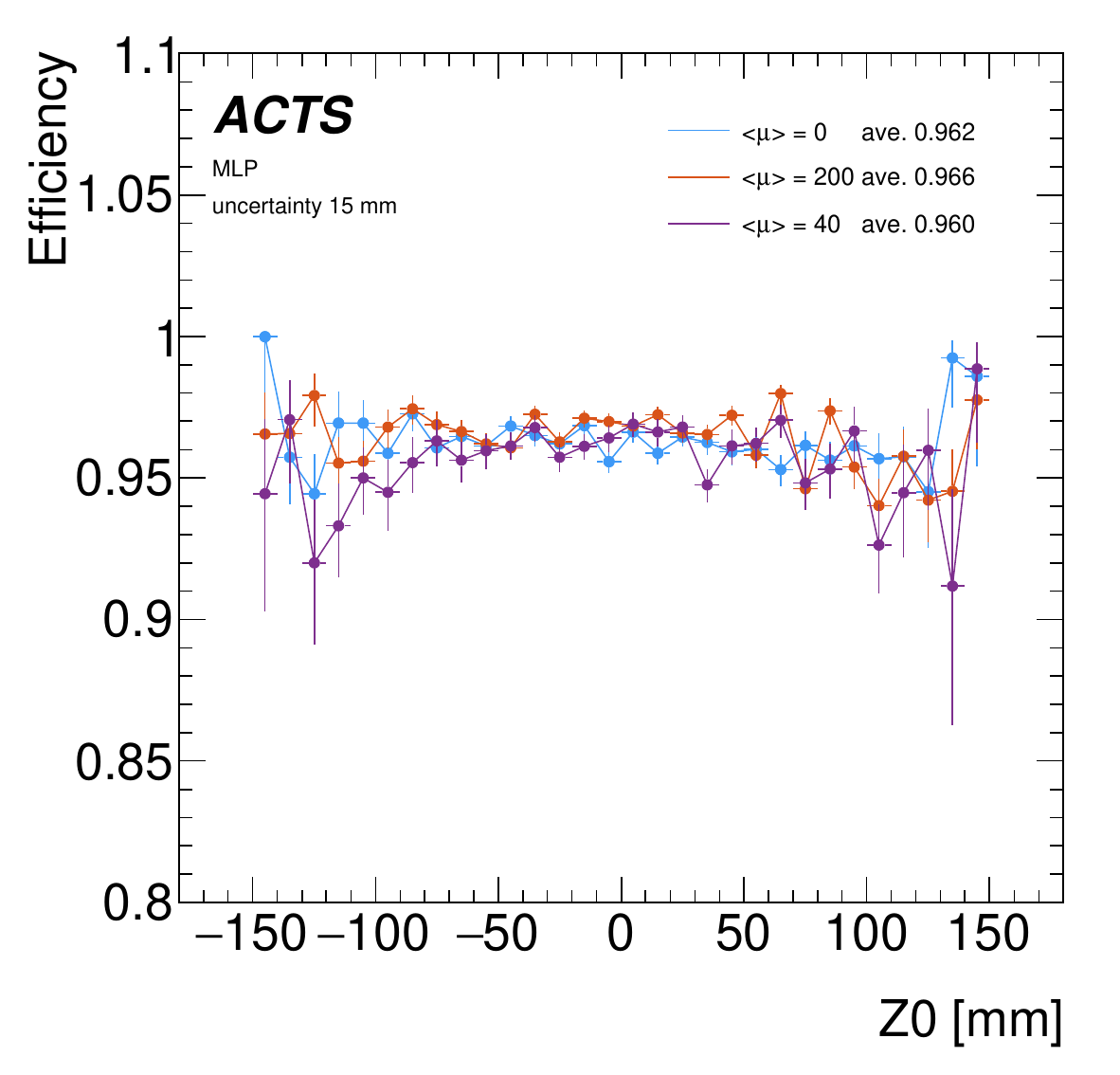}

    \caption{Reconstruction efficiency overlay for different pileup conditions using an MLP NN for the extrapolation as a function of various track parameters. Only tracks with $|\eta|<0.7$ are considered. The true and extrapolated tracks are required to have a minimum of 8 hits. An extrapolated track is considered "matched" to a true track if more than 50\% of the hits in the extrapolated track, excluding seed hits, stem from the same particle. The reconstruction efficiency is calculated as the fraction of truth tracks that are matched to at least one extrapolated track within an uncertainty window of 20mm.}
    \label{fig:efficiencypileup}
\end{figure}

The performance of the NN are evaluated using tracks that are contained in the barrel region of the detector, which is defined as $|\eta|<0.7$. Both the true and extrapolated tracks are require to have a minimum of 8 hits, ensuring that they have enough energy to traverse the detector. An extrapolated track is considered "matched" to a true track if more than 50\% of the hits in the extrapolated track, excluding seed hits, stem from the same particle, as every track seed is comprised of hits belonging to the true track. The NN extrapolator's reconstruction efficiency is calculated as the fraction of truth tracks that are matched to at least one extrapolated track: 
\begin{equation}
    \epsilon_{\text{extrap}}=\frac{N_{matched}}{N_{true}}.
    \label{eqn:efficiency}
\end{equation}
As the current performance is calculated using truth seeds, the overall tracking efficiency of full algorithm would be $\epsilon_{\text{seed}} \times \epsilon_{\text{extrap}}$.  A few examples of matched and unmatched tracks are shown in Figure \ref{fig:tracksdetector}. The reconstruction efficiency as a function of various track parameters for the different NN algorithm is shown in Figure \ref{fig:efficiencyoverlay20}. As observed in Table \ref{table:modelCompare}, due to the largest expected residual between the true and predicted hits, the ESN NN has the lowest efficiency. The other three networks have a similar efficiency when the tracks originate close to the interaction point and low $d_{0}$ parameter. The MLP and RNN algorithm have lower performance at higher $d_{0}$ values, with the MLP NN showing a larger sensitivity. The pileup dependency of the MLP NN is showing in Figure \ref{fig:efficiencypileup}. Across all variables, this dependency is found to be negligible.

\section{Track Overlap removal and fake rejection with NNs}
\label{sec:overlap}
This section describes the Neural Network built to reduce the fake tracks after the pattern recognition step. A similar strategy as the one described in Ref. \cite{ATLASTDR} is used. 

The model architecture used for this step is a densely connected MLP with 2 layers of 32 nodes each. The input features used are the 10 hits coordinates. If a track consists of less than 10 hits, the input features are padded with zero. The output of this NN is a single score that is related to the probability of the candidate track originating from a true particle. The final resulting architecture of this model is 30 x 32 x 32 x 1.

The training dataset used for this NN is based on the output of the MLP extrapolator described in Section \ref{sec:trackseeding}. True labelled tracks are classified as combination of hits where more than 95\% of the hits originate from the same truth particle. Similarly, the fake labelled tracks are hits combination where less than 50\% of the hits originate from one truth particle. Mean squared error is used as the loss function. 

\begin{figure}[h]
\centering
\includegraphics[width=0.45\textwidth]{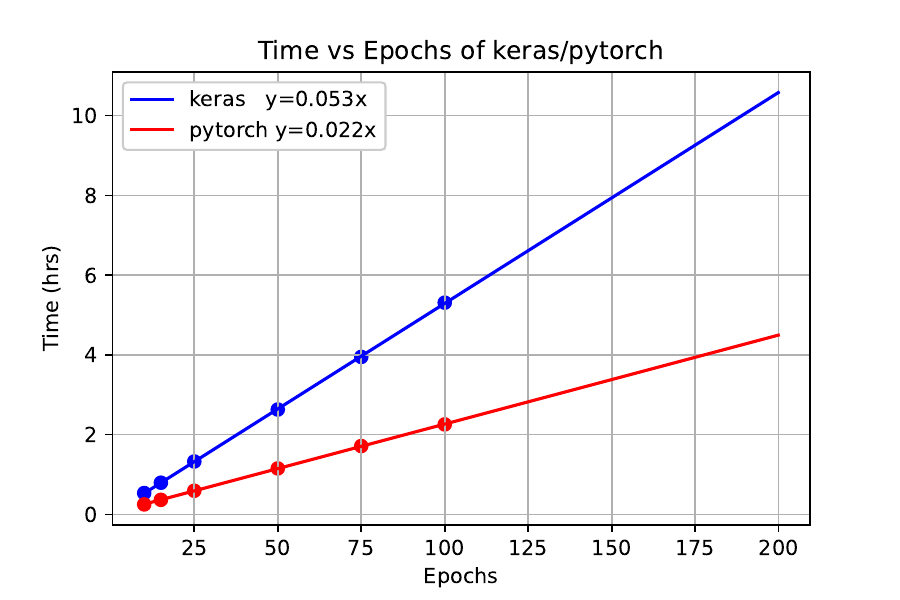}
\includegraphics[width=0.45\textwidth]{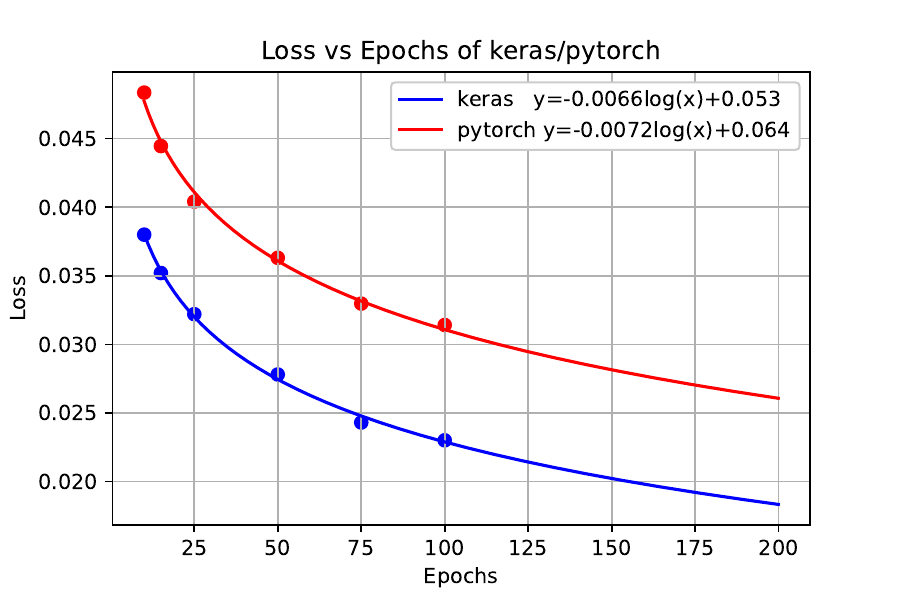}
\caption{The loss and time required as a function of the training epoch for the neural network designed to perform fake track rejection and overlap removal. Implementation using Keras and Pytorch libraries are compared. }
\label{fig:PytorchKerasCompare}
\end{figure}

\begin{figure}[h]
\centering
\includegraphics[width=0.5\textwidth]{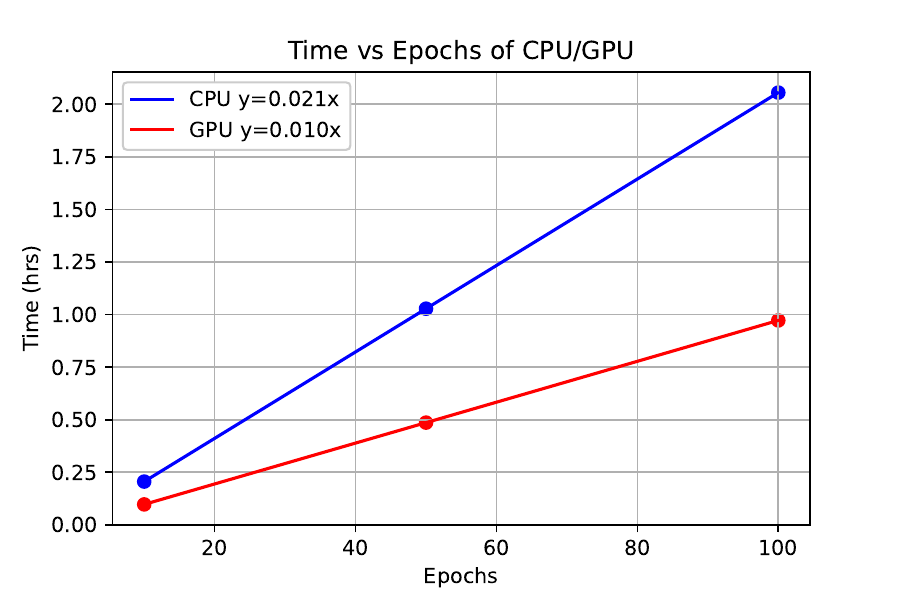}
\caption{The training time as a function of the number of epochs for the neural network designed to perform fake track rejection and overlap removal. Training using CPUs and GPUs are compared. }
\label{fig:GPUCPUCompare}
\end{figure}

The implementation and training of this network was performed using both the Keras \cite{chollet2015keras} and Pytorch \cite{NEURIPS2019_9015} library with nominal optimization of the training hyper-parameters. As shown in Figure~ \ref{fig:PytorchKerasCompare}, training time required for the Pytorch implementation is typically shorter than the Keras implementation. However, Keras implementation has a lower loss value for the same number of training epochs. Similarly, a comparison of the training using CPUs and GPUs was performed. As shown in Fig. \ref{fig:GPUCPUCompare}, training using GPUs typically required a factor of 2 less time, limited by the access to the samples of data. 

The final implementation of NN is done using the Keras library and the training is performed on GPU. Additionally it was observed the performance of the NN was significantly improved  by:
\begin{enumerate}
    \item Removing the rotational symmetry through rotating the input track candidate such that the first hit is at $\phi = 0$,
    \item Scaling the hit coordinates such that max position is O(1),
    \item and finally ordering the hits with respect to the distance from the center of the detector.
\end{enumerate}
The pre-processing steps are found to improve the NN performance irrespective of the algorithm used to create the hit combinations.

\subsection*{Application and Performance}
\begin{figure}[h]
    \centering
\includegraphics[width=0.55\textwidth]{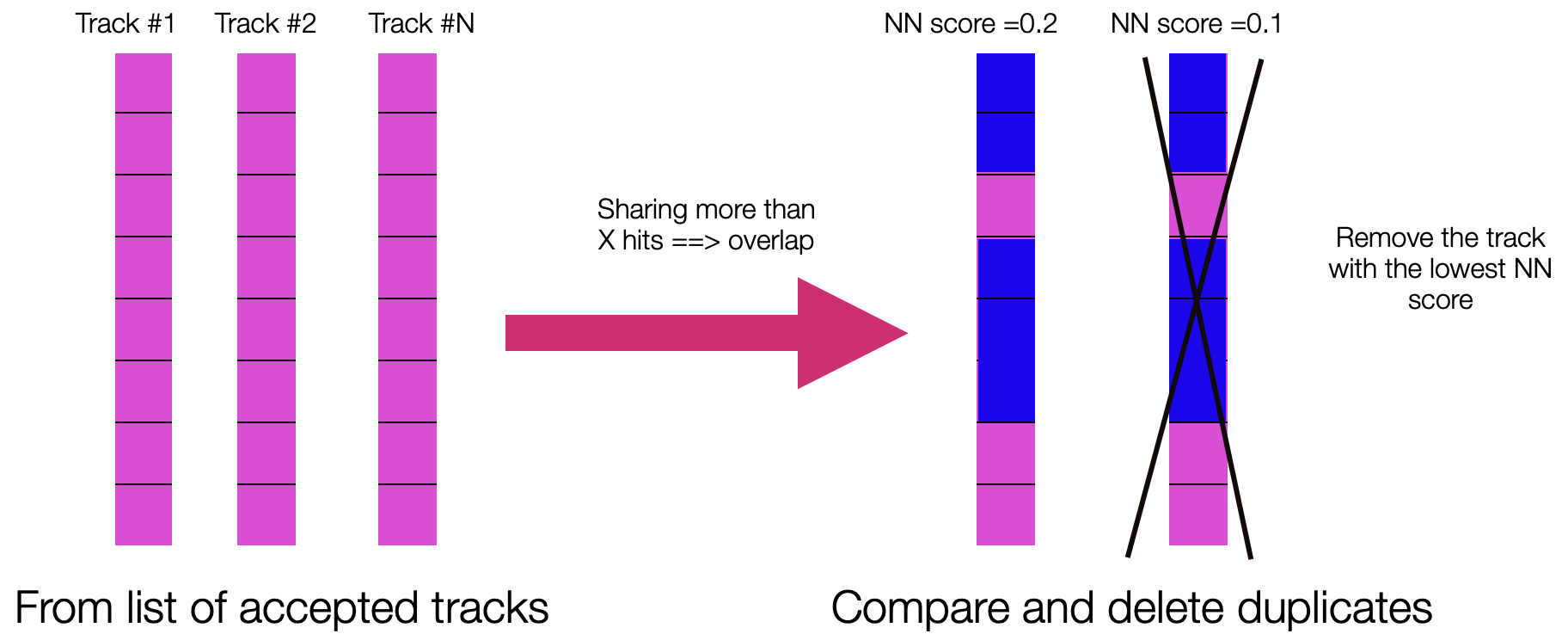}
\includegraphics[width=0.4\textwidth]{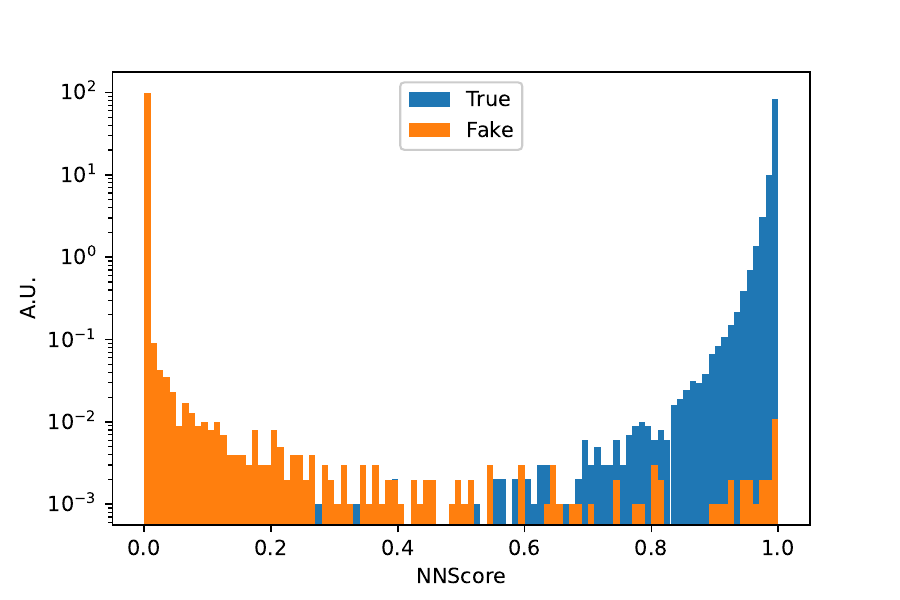}
\caption{Left: Sketch of the overlap removal and fake rejection algorithm. If tracks share more than the specific numbers of hits, only the track with the higher NN score is kept. Right: Distribution of NN score for truth labelled and fake labelled tracks.
 \label{fig:OverlapRemoval}}
\end{figure}

The overlap and fake rejection setup is similarly applied as the `Hit-Warrior' algorithm described in Ref. \cite{ATLASTDR}. In this step, as outlined in Fig. \ref{fig:OverlapRemoval}, if two or more tracks share N hits only the track with the highest NN score is kept to reduce overlapping tracks. Additionally, all tracks are required to have a minimum NN score to reject fake tracks. 

\begin{figure}[h]
    \centering
\includegraphics[width=0.45\textwidth]{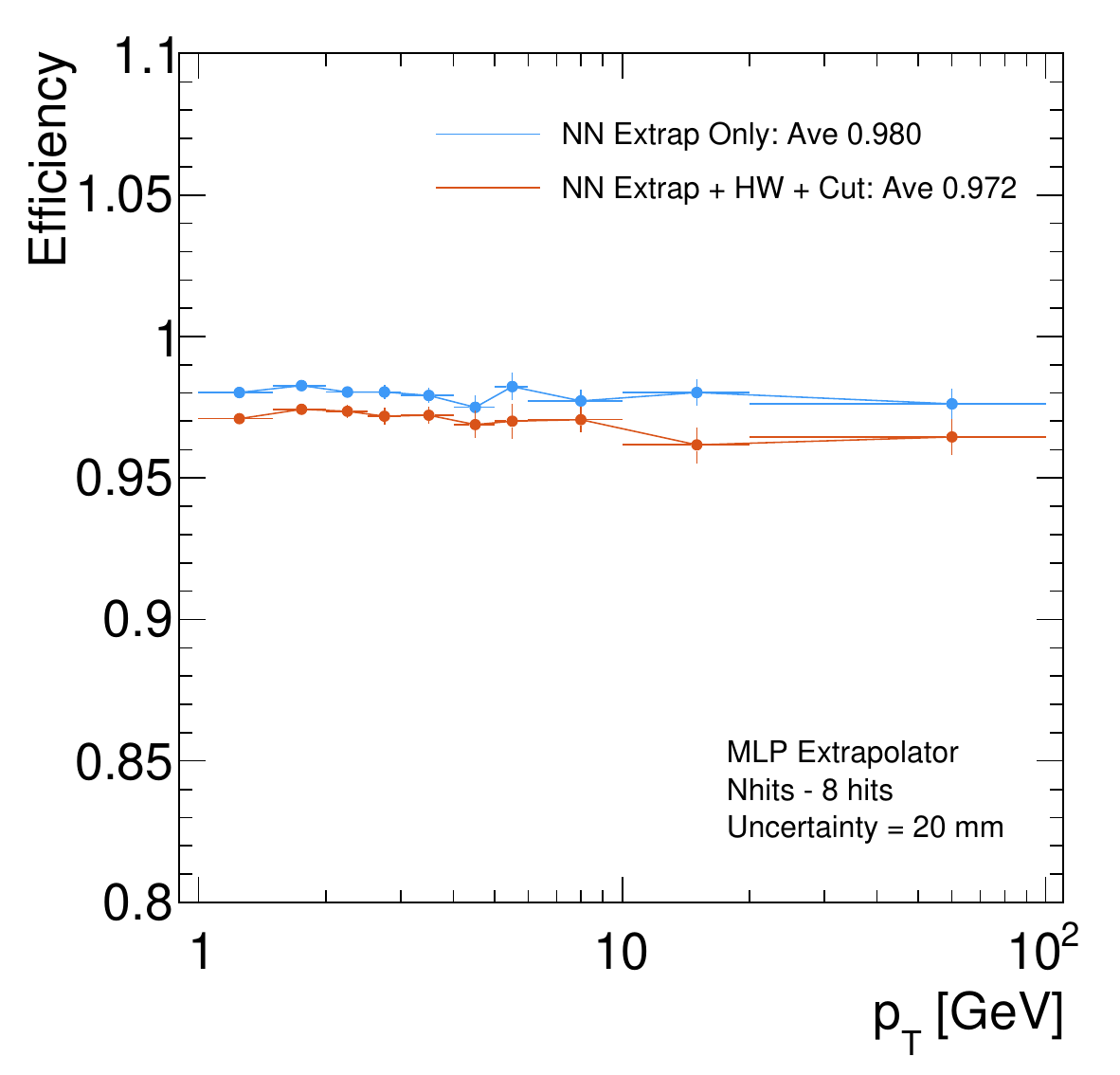}
\includegraphics[width=0.45\textwidth]{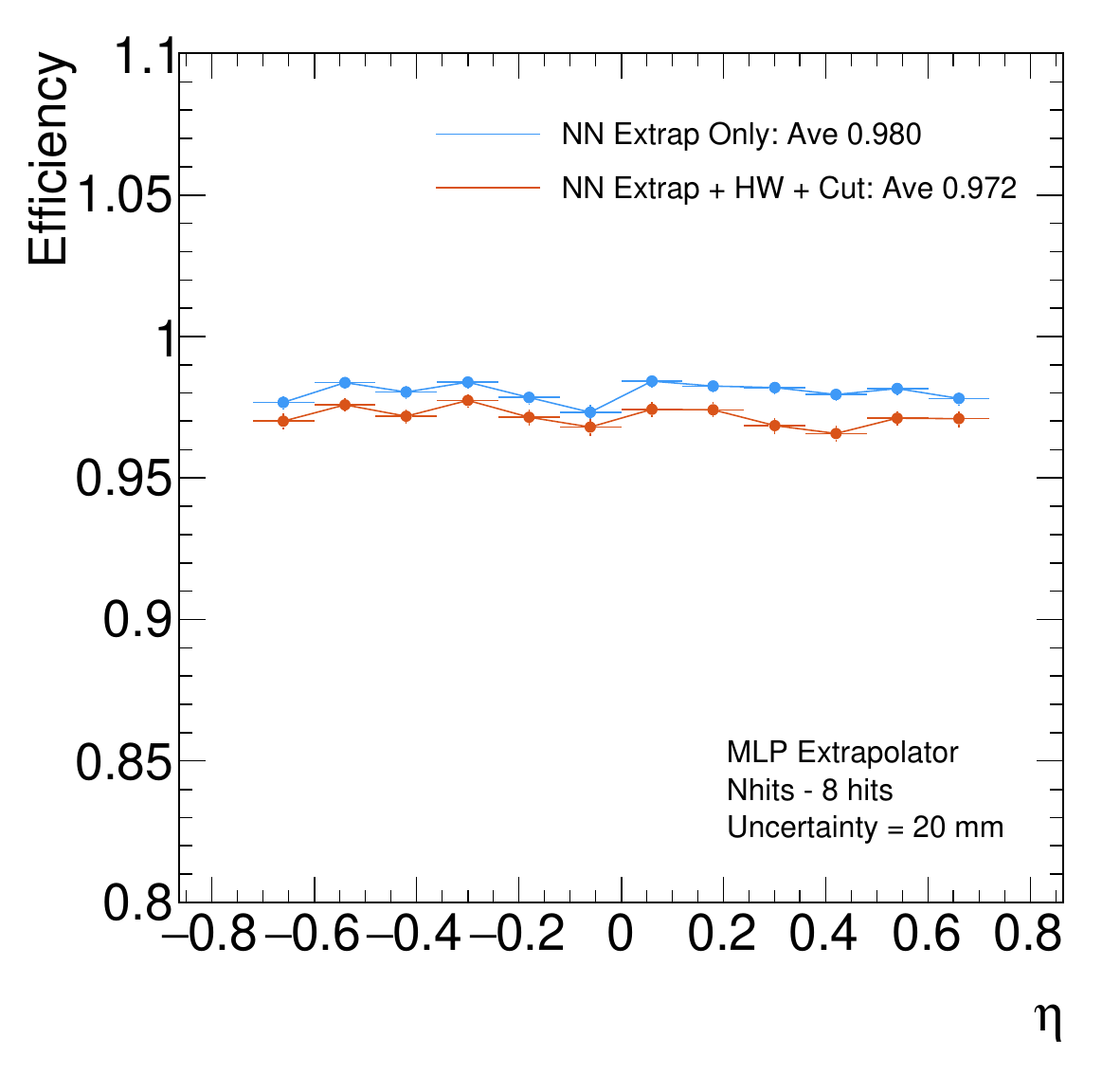}
\includegraphics[width=0.45\textwidth]{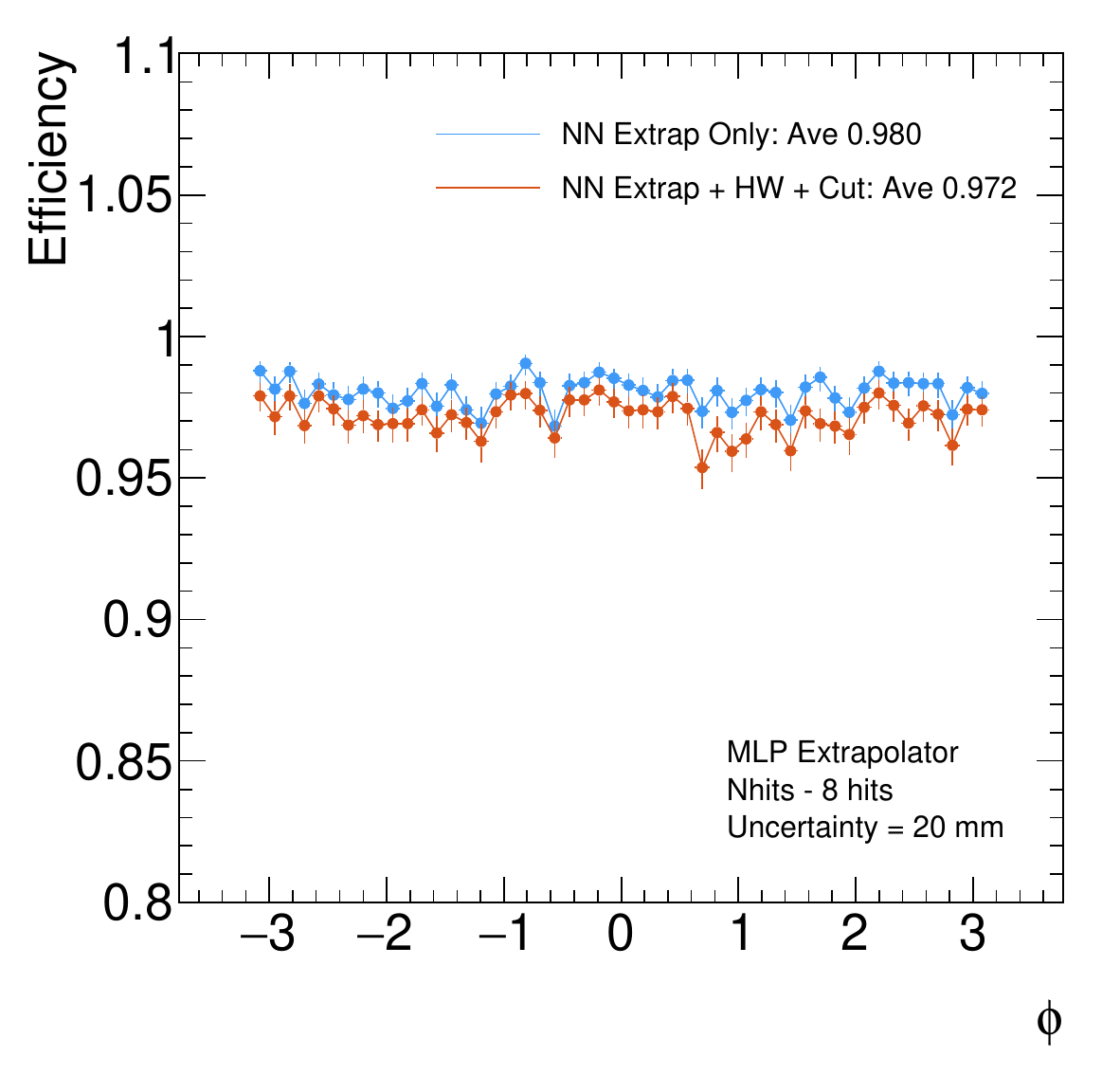}
\includegraphics[width=0.45\textwidth]{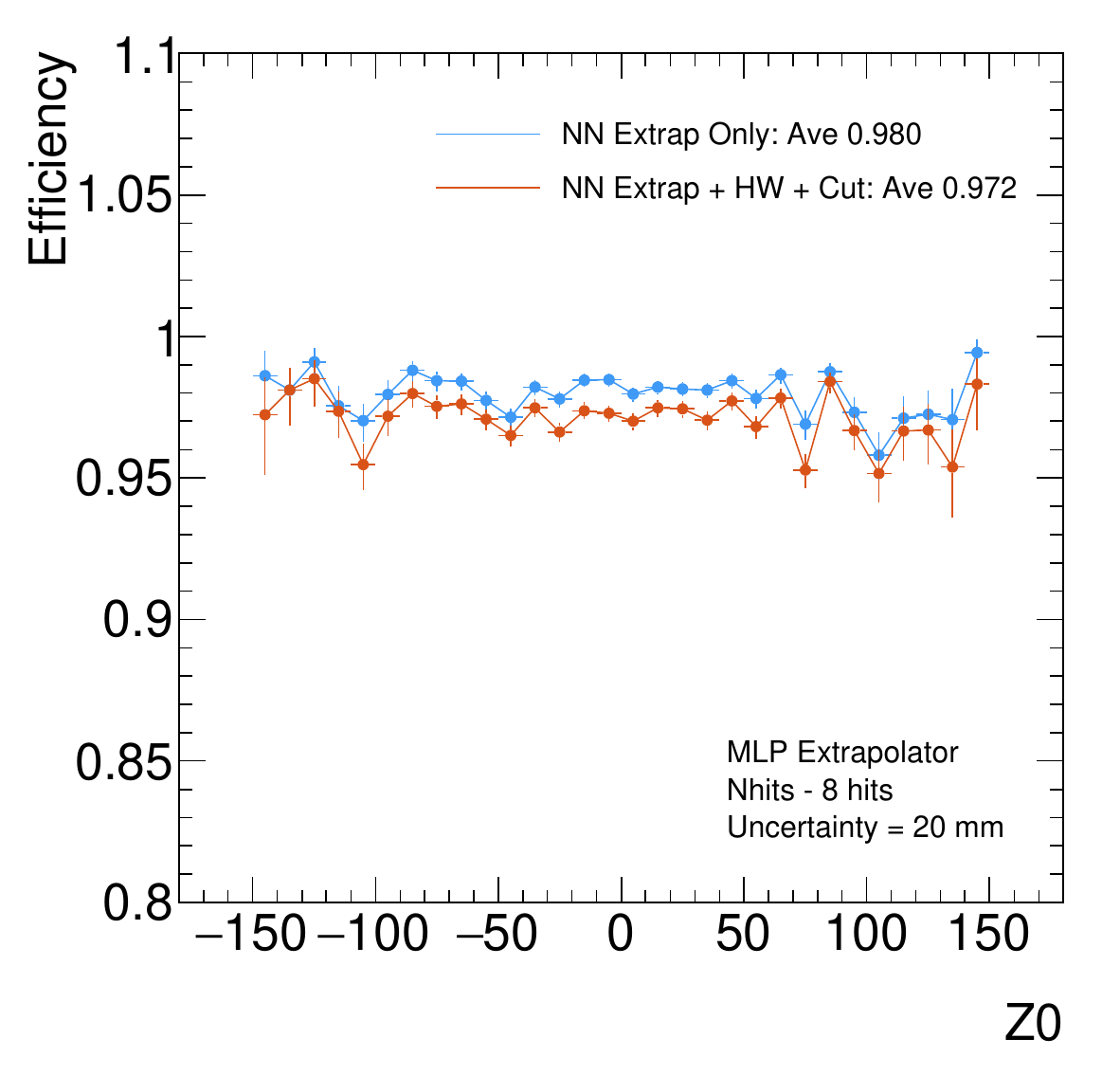}
\includegraphics[width=0.45\textwidth]{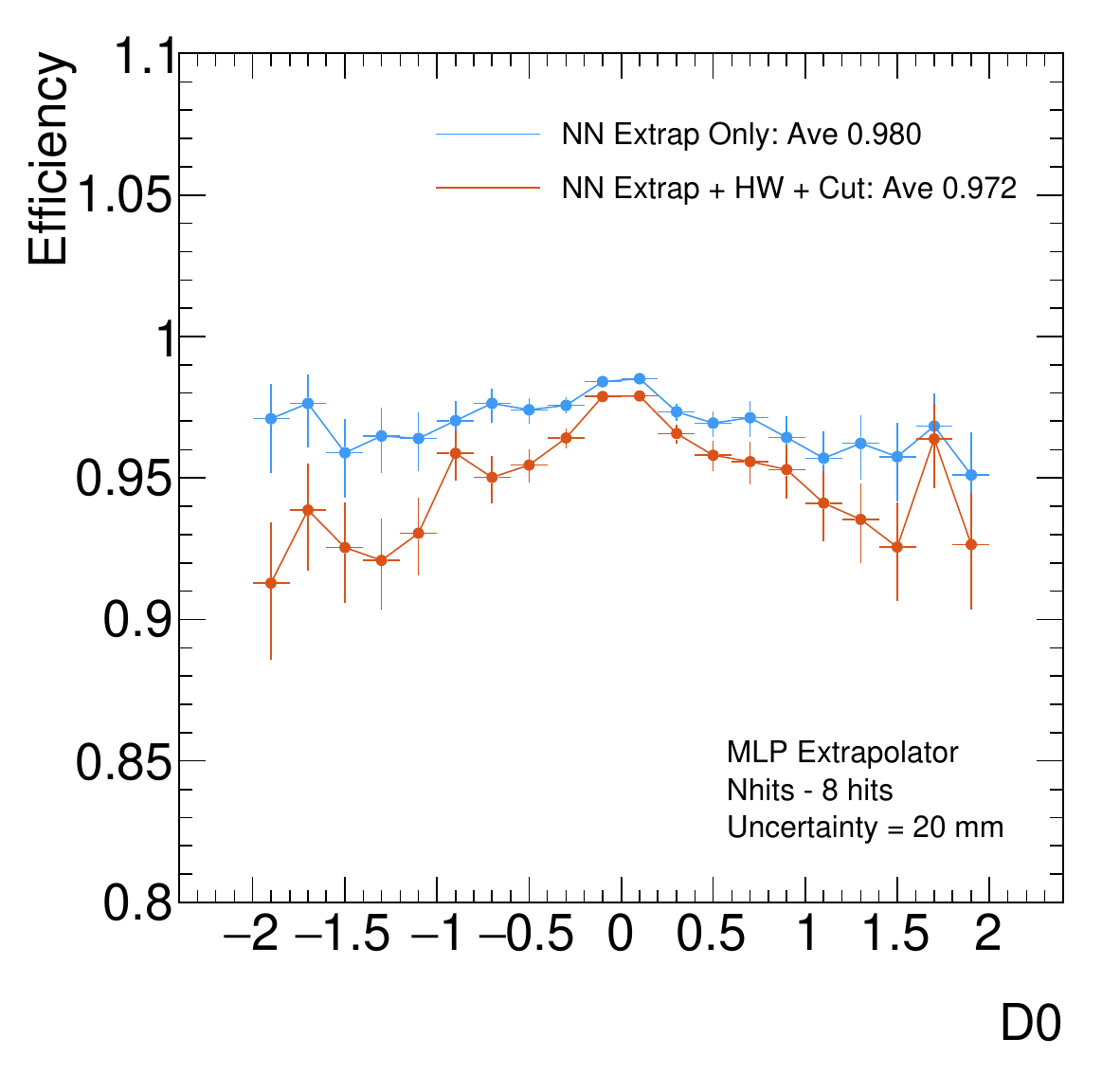}
\includegraphics[width=0.45\textwidth]{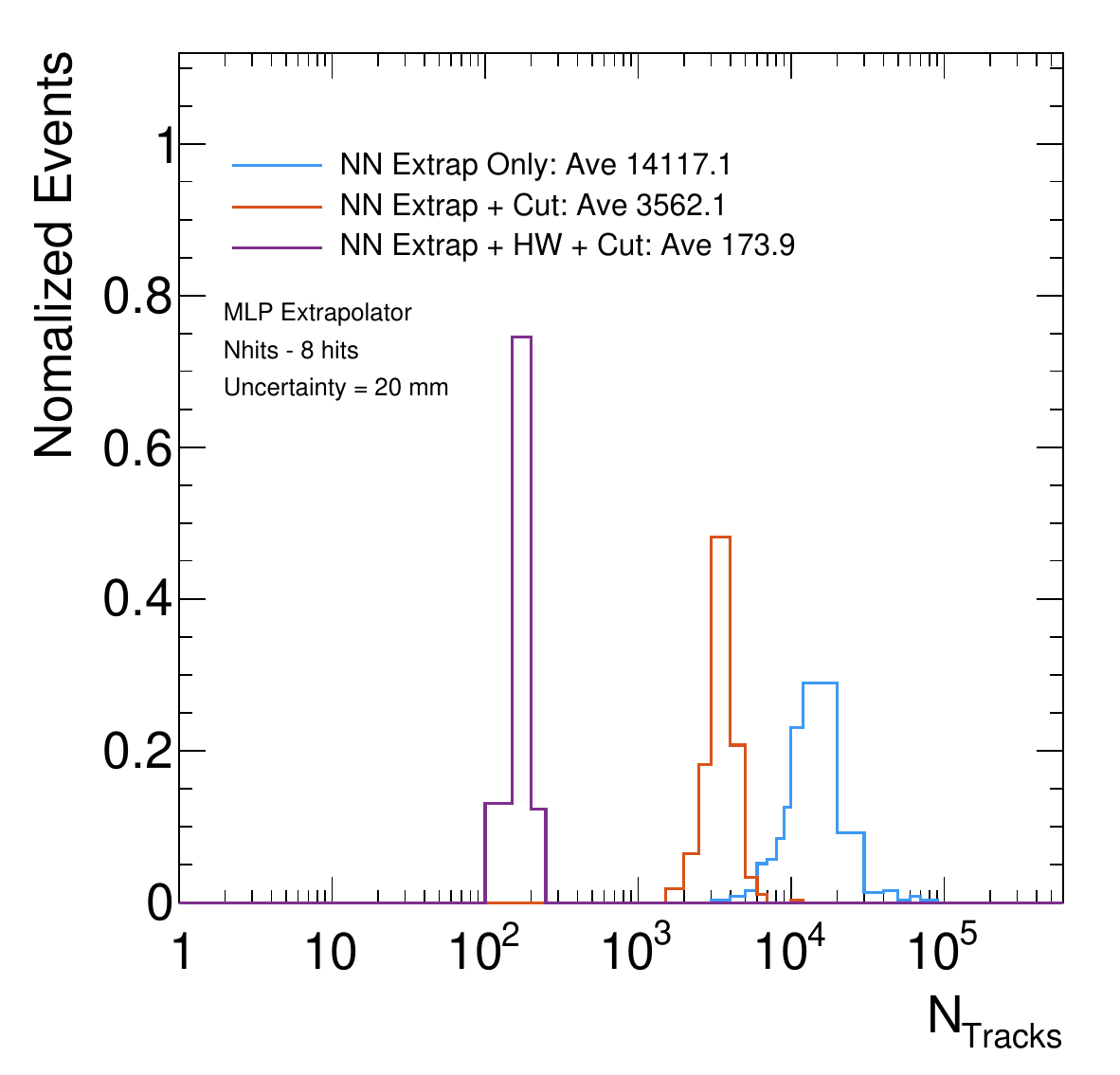}
\caption{Track efficiency as a function of the track parameters for true tracks and the average number of tracks reconstructed by the MLP extrapolation for a window size of 20 mm for $<\mu> = 200$. The Efficiency is compared after the application of the overlap and fake removal NN step.  \label{fig:OverlapPref}}
\end{figure}

The output of the training NN for truth and fake labelled tracks is shown in Fig. \ref{fig:OverlapRemoval}. The performance of this algorithm with a cut on the NN score being greater than 0.5 and overlap threshold at 8 hits is shown in Fig. \ref{fig:OverlapPref}. After the application of this step, a small loss of 0.9\% is observed in the truth track efficiency. However, an approximately factor of 80 reduction in total tracks is observed. The average purity of the MLP extrapolator increases from 0.9\% to  70\% with the application of this step. 

\section{Estimation of resources on FPGAs}
\label{sec:resources}
To translate the NNs for the extrapolator and the overlap and fake removal step for inference on FPGA, the HLS4MLv0.6 library is used \cite{hls4ml}. This package provides a convenient way to implement these algorithms using high level synthesis. The MLP based extrapolator is used to estimate the resource usage. The Xilinx Alveo U250 FGPA is targeted for the implementation. 

\begin{figure}[h]
    \centering
\includegraphics[width=0.45\textwidth]{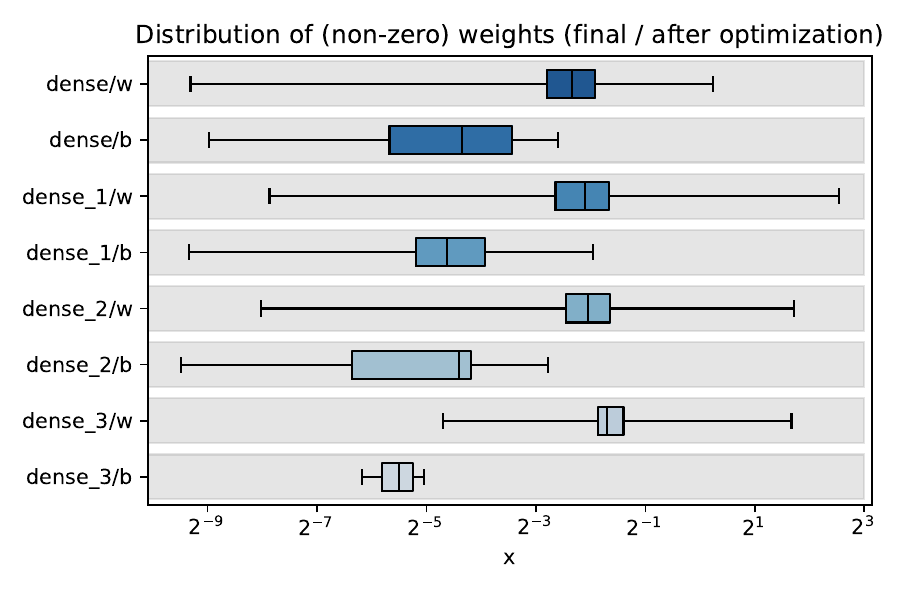} 
\includegraphics[width=0.45\textwidth]{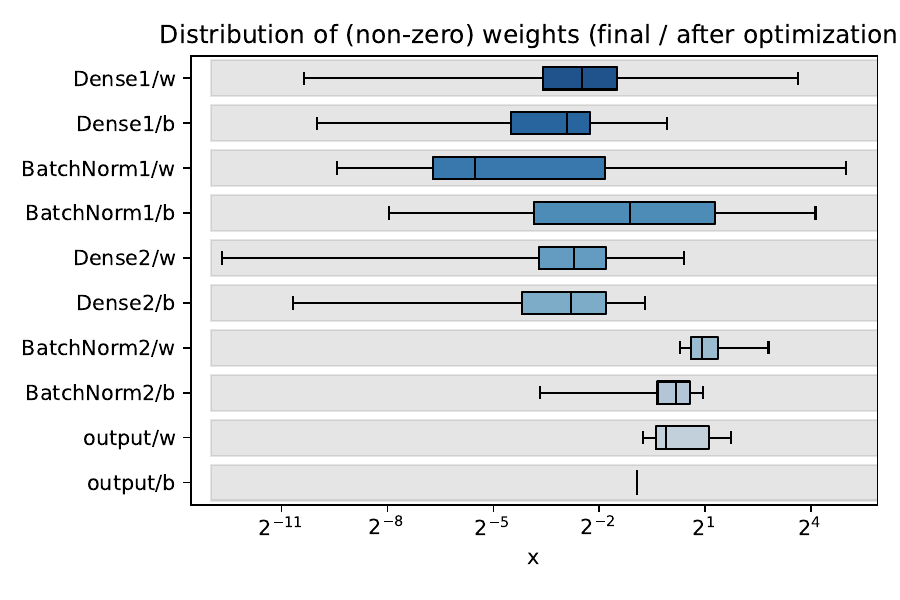}
\caption{Distributions of weights and bias for each layer of the extrapolator and overlap and fake rejection NN. \label{fig:profileWeight}}
\end{figure}

\begin{figure}[h]
    \centering
\includegraphics[width=0.45\textwidth]{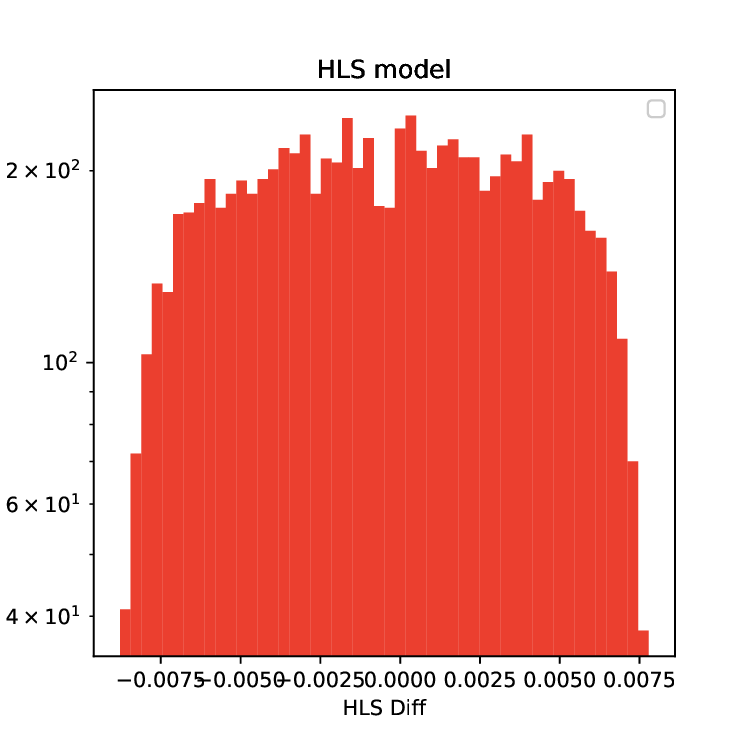} 
\includegraphics[width=0.45\textwidth]{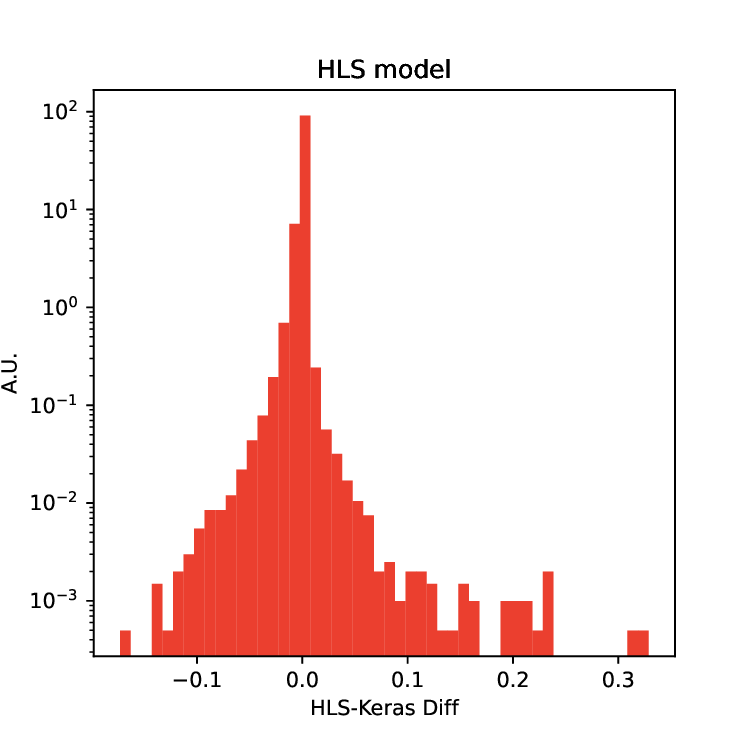}
\caption{Numerical difference in the output predictions between the original NN and the ones quantized for implementation on the FPGA for the MLP extrapolator and the overlap and fake rejection step. \label{fig:NumDiff}}
\end{figure}

To reduce the expected resources needed on a FPGA, the NNs were sparsely trained with pruning of small weights.  For the translation, the reuse factor is set to 1. This implementation will provide the setup with the lowest latency with the highest resource usage. For quantization of the weight, each layer was individually optimized, with an example show in Fig \ref{fig:profileWeight} of the typically distributions of weight and the span of the numerical representation of those weight on the FPGA. The difference in the output predictions between the two different representation of the NN are shown in Fig. \ref{fig:NumDiff}. The final resource estimates are provided in Table \ref{tab:ResourceUsage}.

\begin{table}[]
\centering 
    \caption{Latency and resource usage on an Xilinx Alveo U250 FGPA. \label{tab:ResourceUsage}}
\begin{tabular}{l || c | c c c c}
\toprule
NN           & Latency (ns) & LUT (\%) & FF (\%)  & BRAM/URAM (\%) & DSP (\%) \\
\midrule
Extrapolator              & 50     & 7.       & 0.5                   & < 0.01    & 21 \\[0.2cm] 
Overlap and               & \multirow{2}{*}{50}     & \multirow{2}{*}{18}.      & \multirow{2}{*}{1.}                    & \multirow{2}{*}{< 0.01}    & \multirow{2}{*}{31} \\ 
Fake removal              &        &          &                       &           &    \\ 
\bottomrule

\end{tabular}
\end{table}
\FloatBarrier

\section{Conclusion}

In this paper we have presented a novel ML based approach to charged particle tracking in a generic detector geometry. A set of neural networks of various architectures are trained on particle tracks, using spatial per hit information as input to the networks. In most cases, the NNs are able to precisely approximate the dynamics of a charged particle traversing through a magnetic field within the barrel of a cylindrical particle detector. The approximated function is used to predict the hit coordinates along a track, and hits are then combinatorially matched to the predicted hit location. The overlap removal algorithm is applied to the set of produced tracks, and a second neural network prunes out fake and duplicate tracks from the remaining set. The track finding efficiency and number of generated tracks resulting from this algorithm are compared with the Hough Transform in <$\mu$>=0,40,200 pileup conditions. In the case of <$\mu$>=200, the number of tracks produced via the ML approach shows a factor 1000 reduction compared to that of the HT, at the cost of only 1\%-2\% in efficiency, which can be tuned according to the needs of the experiments. The resource usage for a NN implemented on an FPGA is estimated, along with the per prediction latency. 

\section{Future Work}
The tracking approach outlined here is promising in both its efficiency and amenability to heterogeneous computing architectures. The next steps in developing this approach include extending predictions to the endcap layers, predicting per hit uncertainties, and optimizing the FPGA implementation of the algorithm.

\FloatBarrier
\pagebreak

\appendix
\section{Hough Transform details}
\label{app:houghtransformdetails}
This appendix describes the full details of the Hough Transform used.

The Hough Transform works by means of a coordinate transform - 
of a hit at $(\rho, \phi)$ in the transverse plane of the tracking detector to 
a line (close to straight) in the $(\phi_0, Aq / \pt)$ plane, 
describing all possible circular tracks through that point and the origin 
where $\phi_0$ is the inclination at the origin and 
$Aq / \pt$ is the track curvature 
(relating to the particle charge $q$ and transverse momentum \pt, and uniform toroidal magnetic field absorbed into the constant $A$). 

Hits are thus grouped into tracks by searching for intersections of these lines, corresponding to track parameters passing through all those points in real space.
This intersection search is performed by drawing the coordinate-transform lines in an ``accumulator'' - a 2D histogram, binned in $\phi_0$ and $Aq / \pt$, and then searching for bins which exceed a threshold value.

\begin{figure}
\centering
\includegraphics[width=0.8\textwidth]{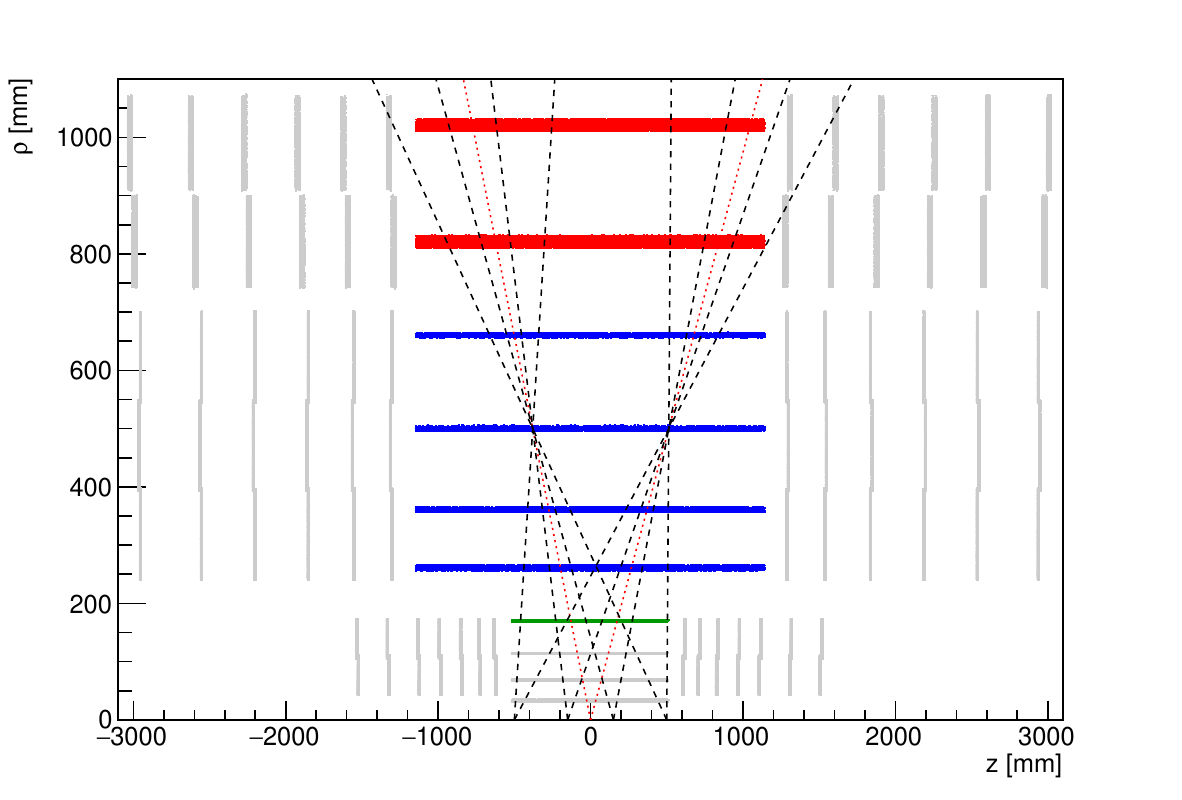}
\caption{ACTS detector layout in the $z - \rho$ plane, with parts not used in the Hough Transform greyed out. The red dashed lines show $\eta=-0.7$ and $\eta=0.9$, and the black dashed lines example trajectories from $v_z = (-500,-140,140,500)$ mm through the $\eta=-0.7$ and $\eta=0.9$ accumulators.}
 \label{fig:ACTSgeo}
\end{figure}

\begin{itemize}
    \item $\phi_0$ range: 0.4 radians wide, plus 0.1 radians overlap on each side $\rightarrow$ 0.6 radians total
    \item $q/\pt$ range: -2 to 2
    \item number of bins per accumulator in $\phi_0$ and $q/\pt$: 500
    \item Detector parts: outer pixel layer, all short strip (4 layers), all long strip (2 double-layers) -- 9 layers total (see Figure~\ref{fig:ACTSgeo})
    \item Hit extension (fill adjacent bins in $\phi_0$): pixel: +2 bins, inner layer of short strip: +1 bin, rest of strip: 0
    \item threshold: require 7 hits
\end{itemize}

\begin{figure}
    \centering
\includegraphics[width=0.45\textwidth]{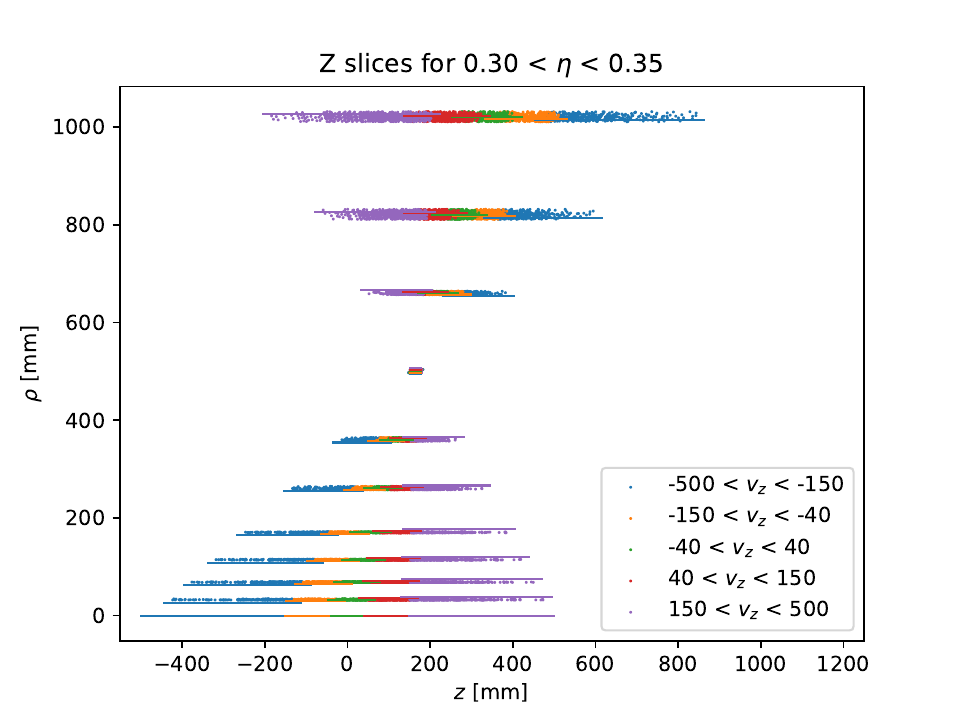}
\includegraphics[width=0.45\textwidth]{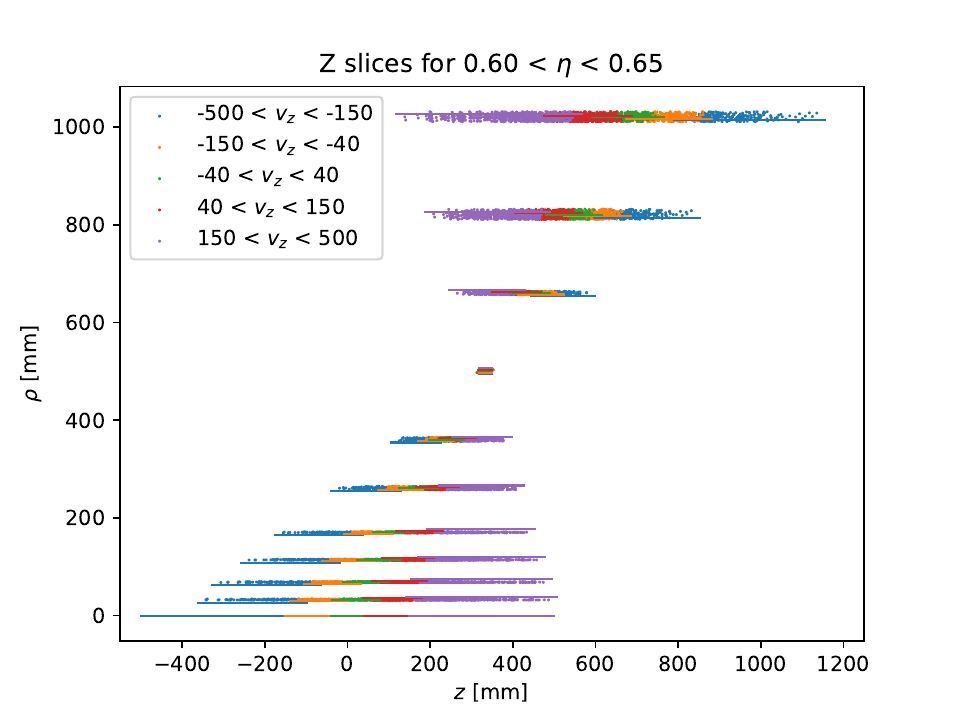}
\caption{Example $z$ slicing - points are hits from events distributed along $|v_z|<500$ mm, lines are the $z$ slices used (offset vertically for legibility). Each $v_z$ slice is defined by an $\eta$ range of 0.05 in the short strip layer at 510~mm, and one of five $v_z$ slices with boundaries at $\pm(40, 140, 500)$~mm, extrapolated to the other detector layers and with a small amount of padding added. There is thus substantial overlap between slices, and many roads are identified multiple times, hence fast overlap removal following the Hough Transform is important.}
 \label{fig:zSlice}
\end{figure}

One accumulator covering the whole detector in $\eta$ is too 
highly occupied for individual tracks to be accurately identified. 
Thus the detector is split up into multiple regions, according to the $v_z$ and $\eta$ of potential track candidates -- an example slicing can be seen in figure~\ref{fig:zSlice}.

\begin{figure}
\centering
\includegraphics[width=0.49\textwidth]{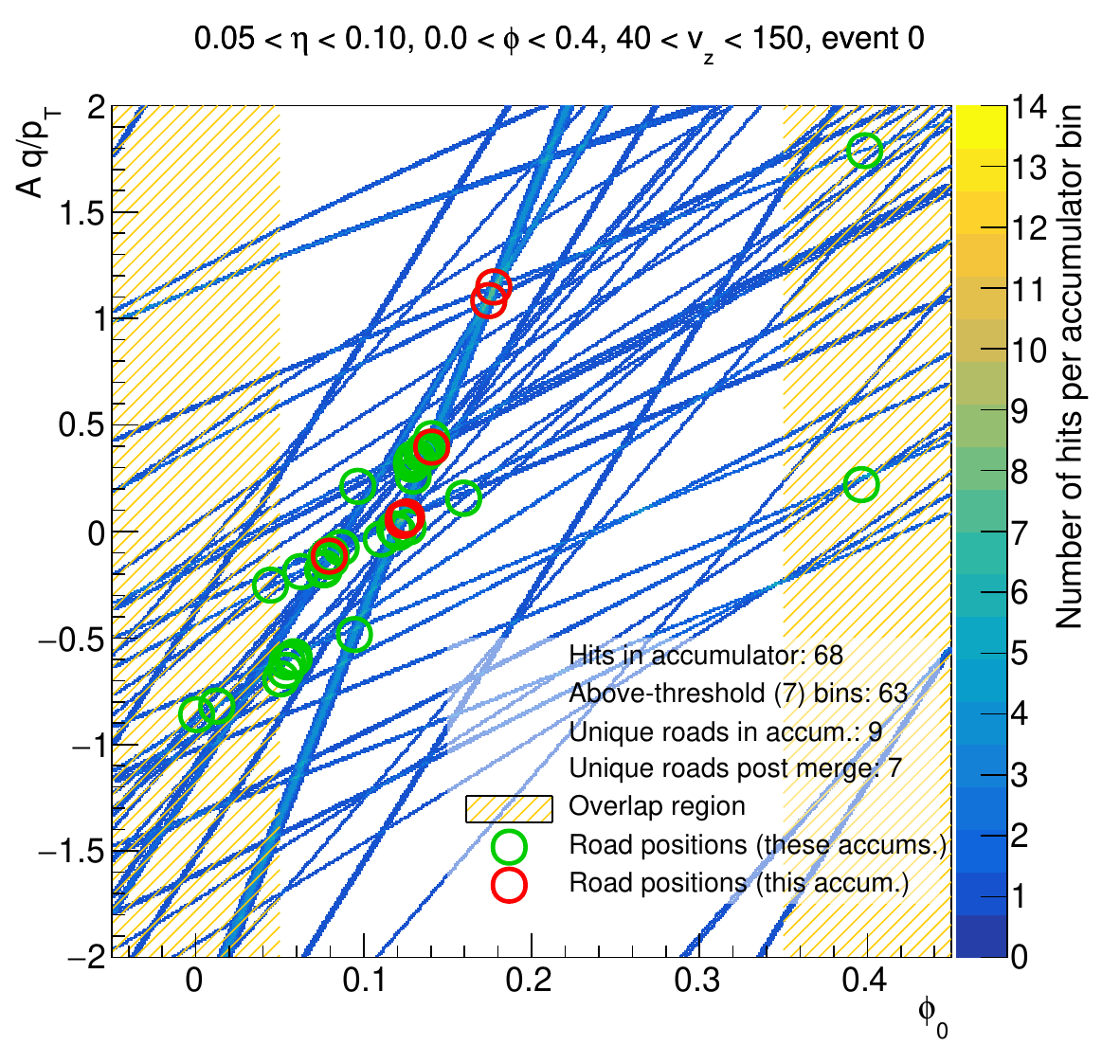}
\includegraphics[width=0.49\textwidth]{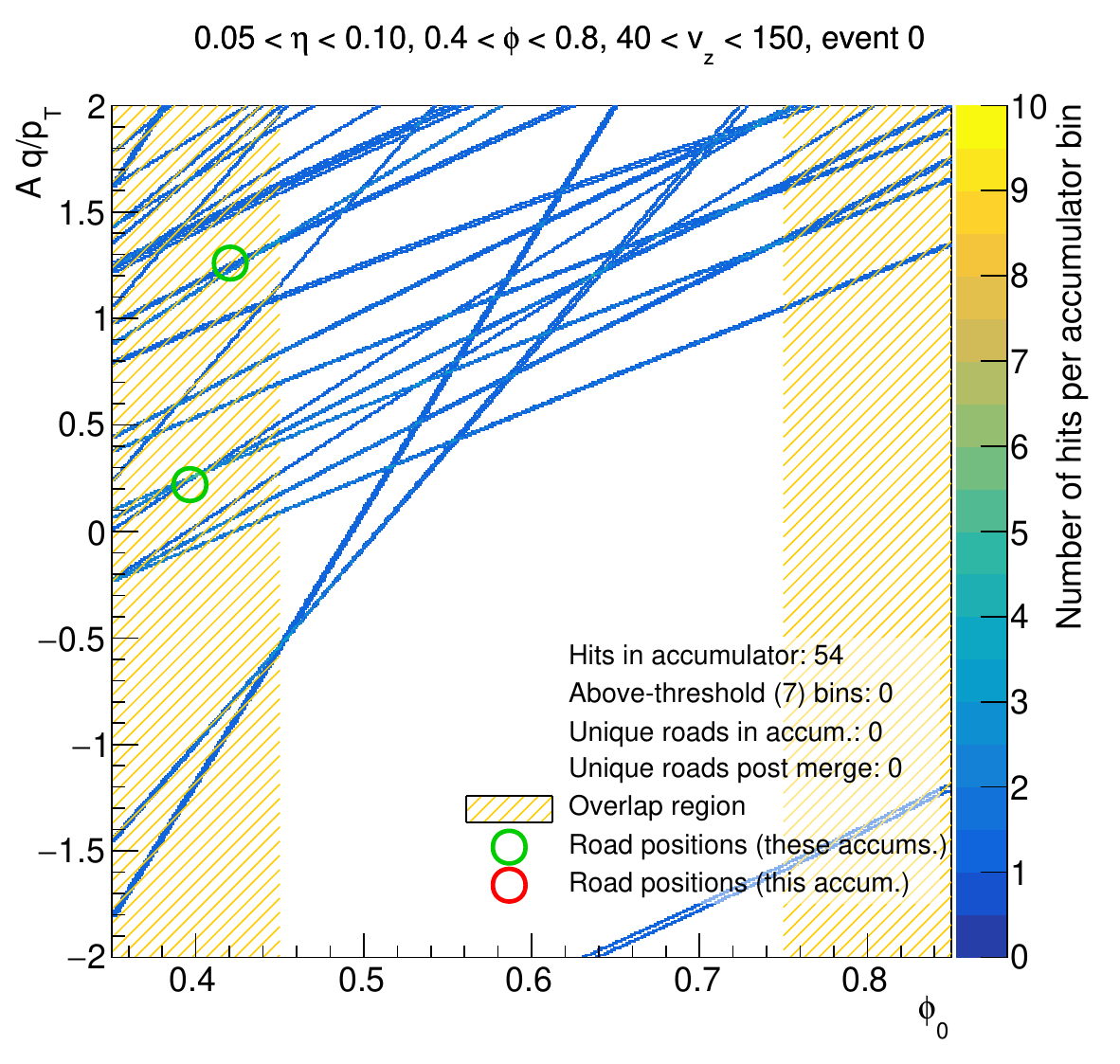}
\includegraphics[width=0.49\textwidth]{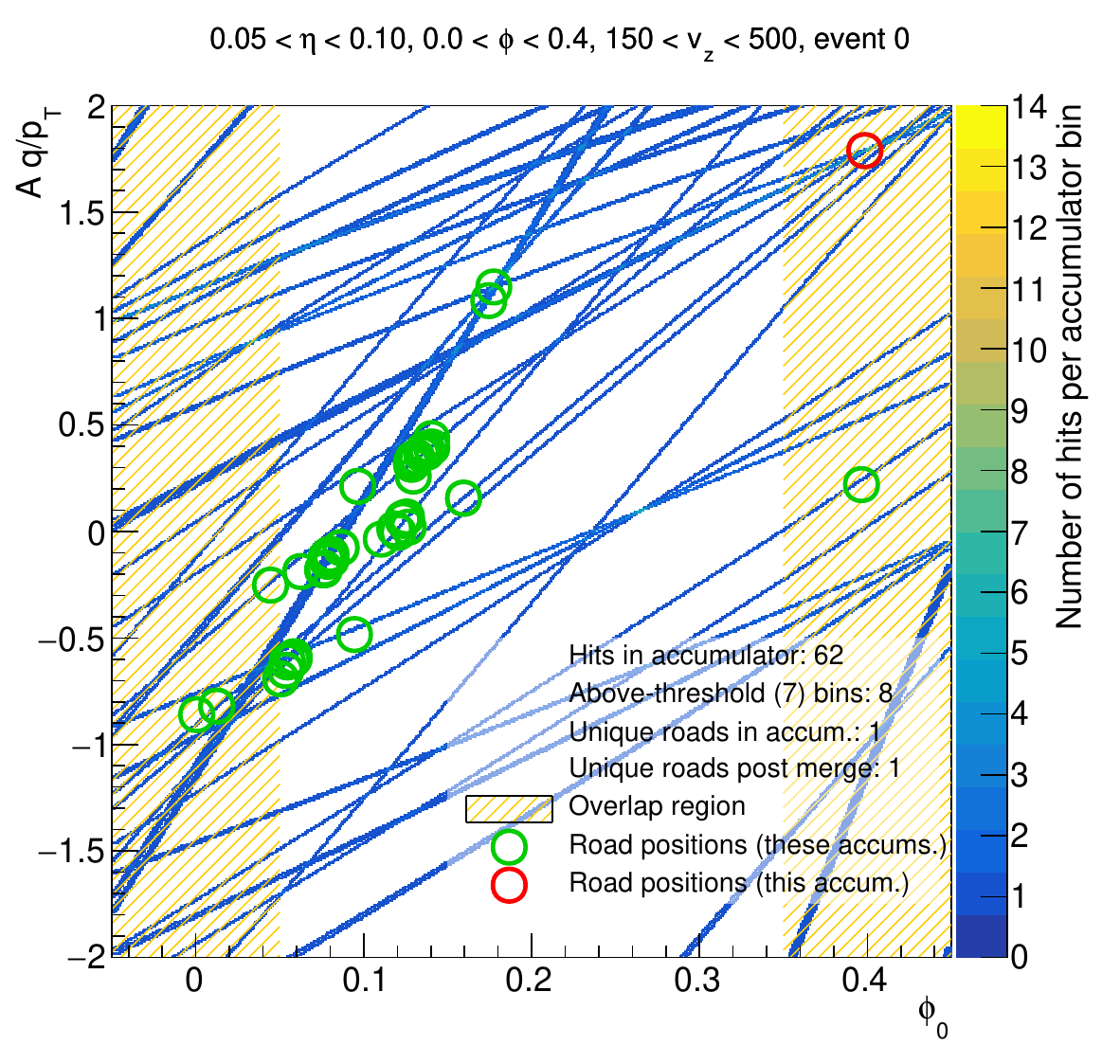}
\includegraphics[width=0.49\textwidth]{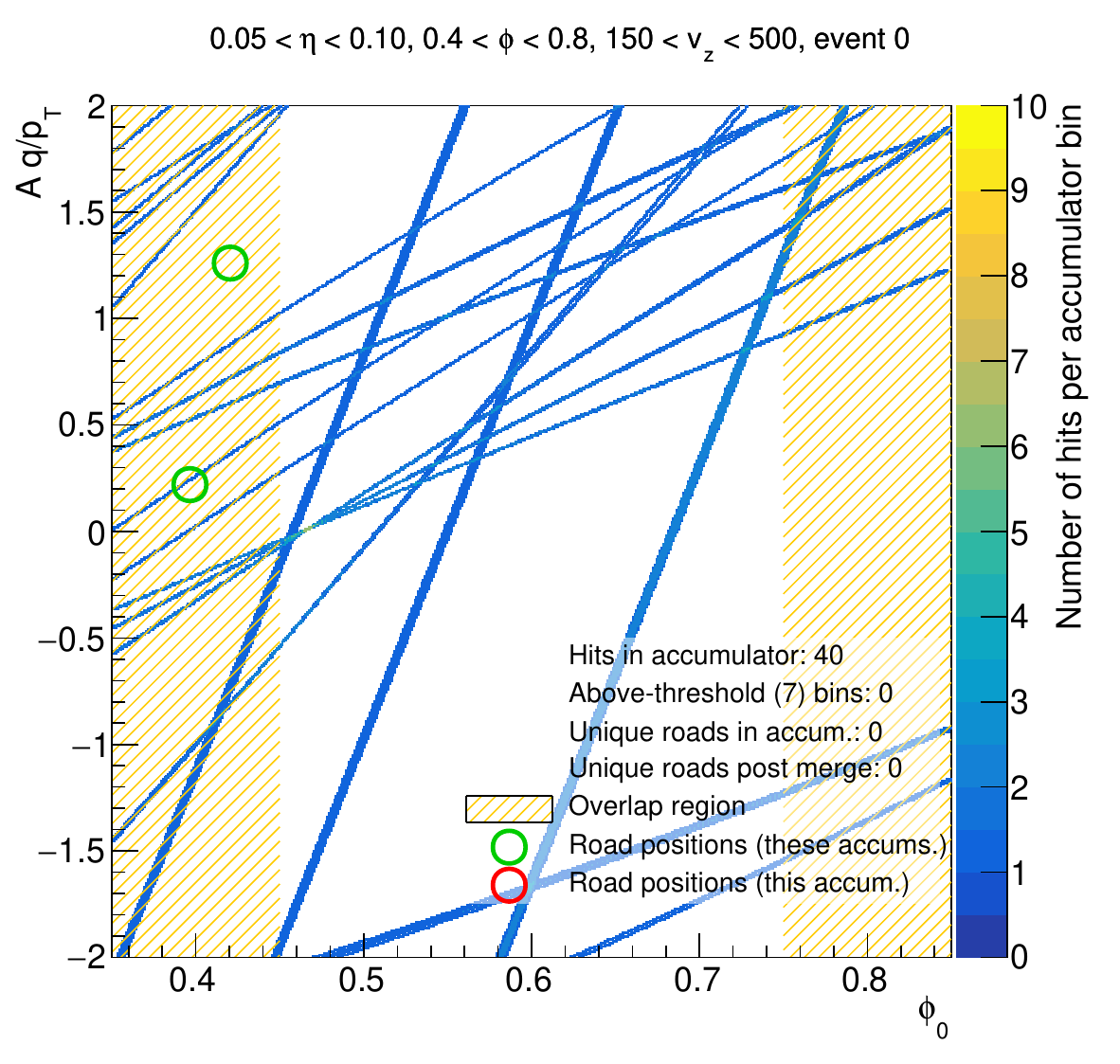}
\caption{Accumulators spanning $0.0 < \phi < 0.8$, for the $z$ slices defined by $0.05 < \eta < 0.10$. Roads present in multiple accumulators are only removed if the $\phi$ and $\eta$ of the accumulator is the same - i.e. between those arranged vertically here, with different $z$-slicing - hence the same roads appear to be present in both sets (left and right).}
 \label{fig:accumulators}
\end{figure}

With this configuration, to cover the full detector in $\phi$ and out to $|\eta|<0.9$ requires 2880 accumulators.
Figure~\ref{fig:accumulators} show ten example accumulator histograms for two $\phi$ segments.

\begin{figure}
\centering
\includegraphics[width=0.49\textwidth]{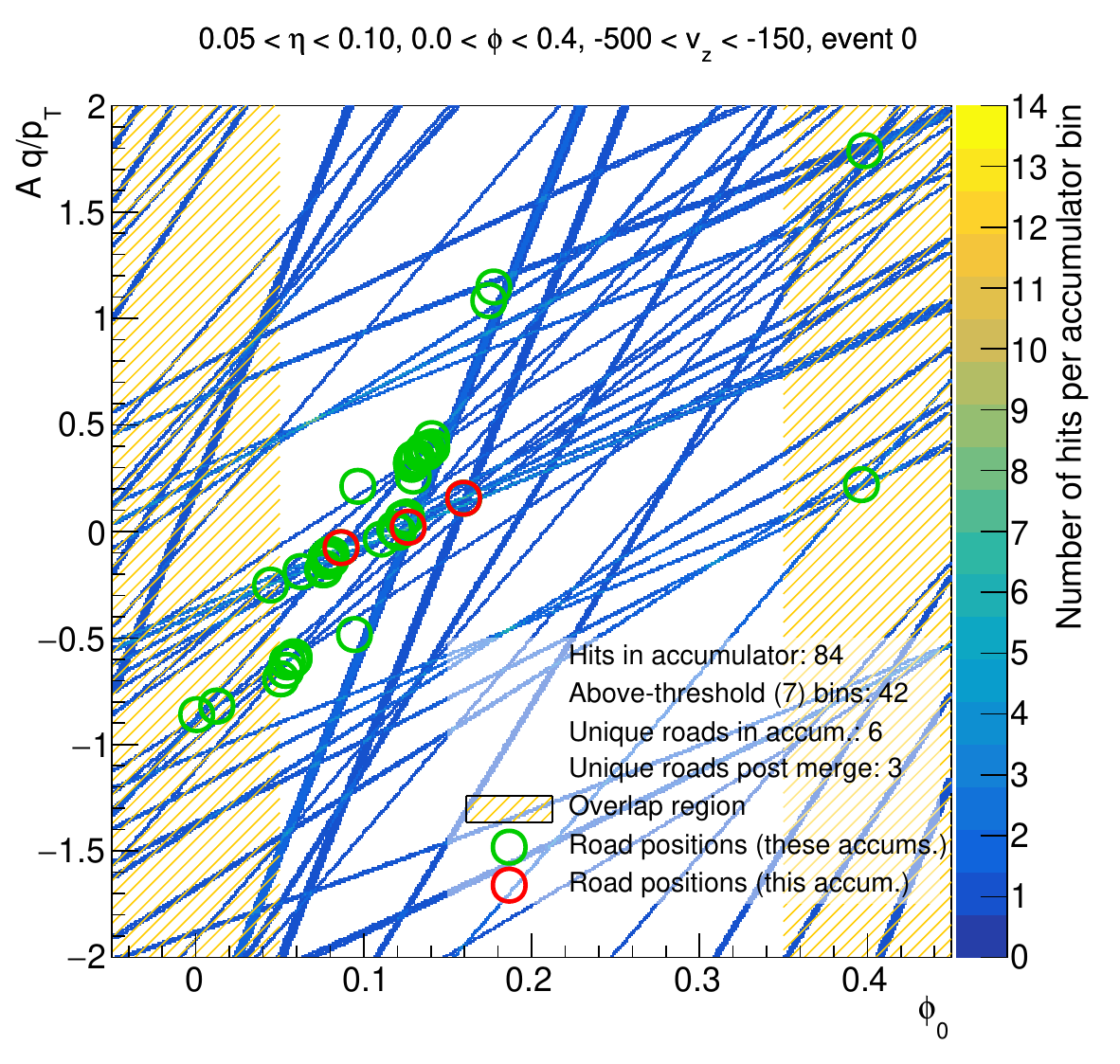}
\includegraphics[width=0.49\textwidth]{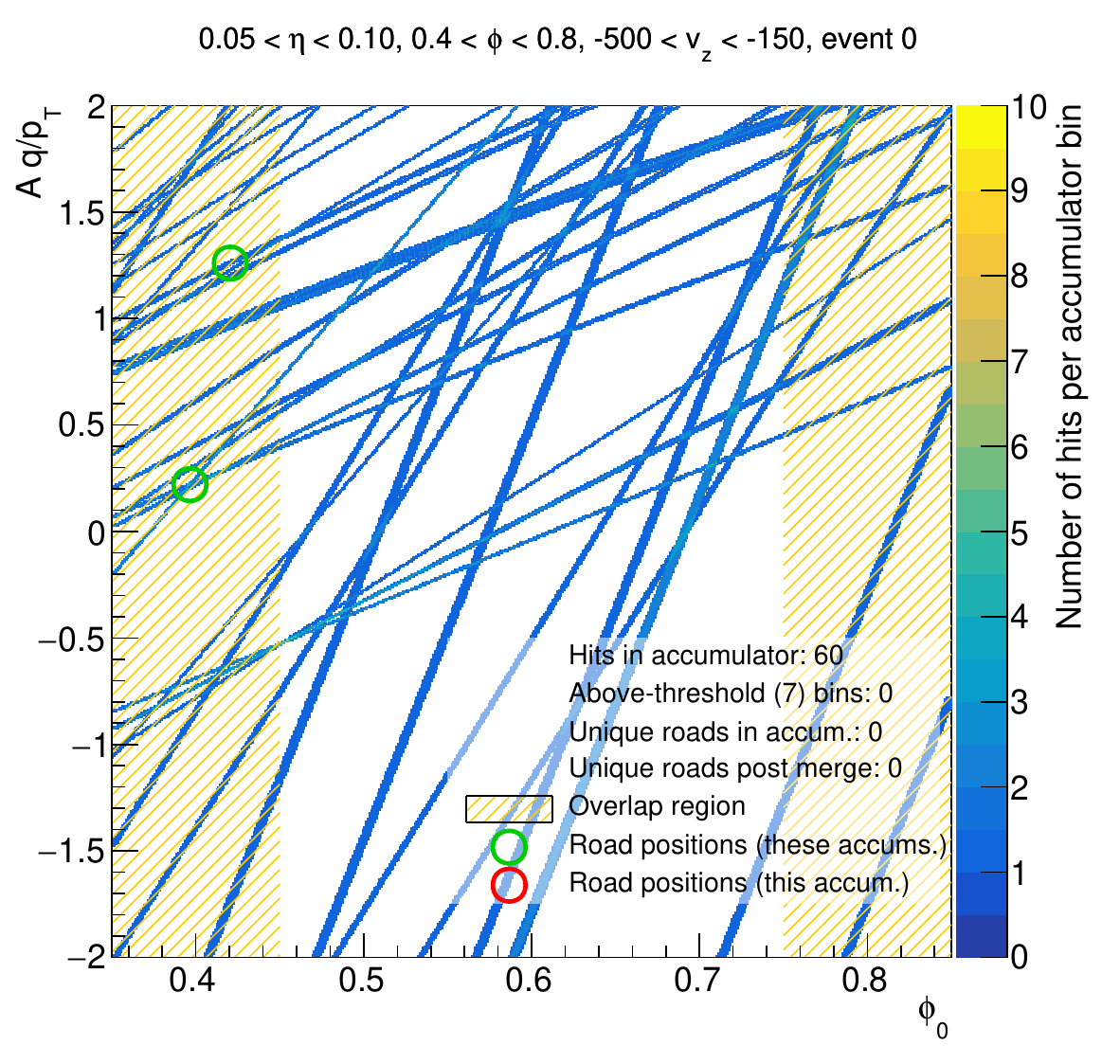}
\includegraphics[width=0.49\textwidth]{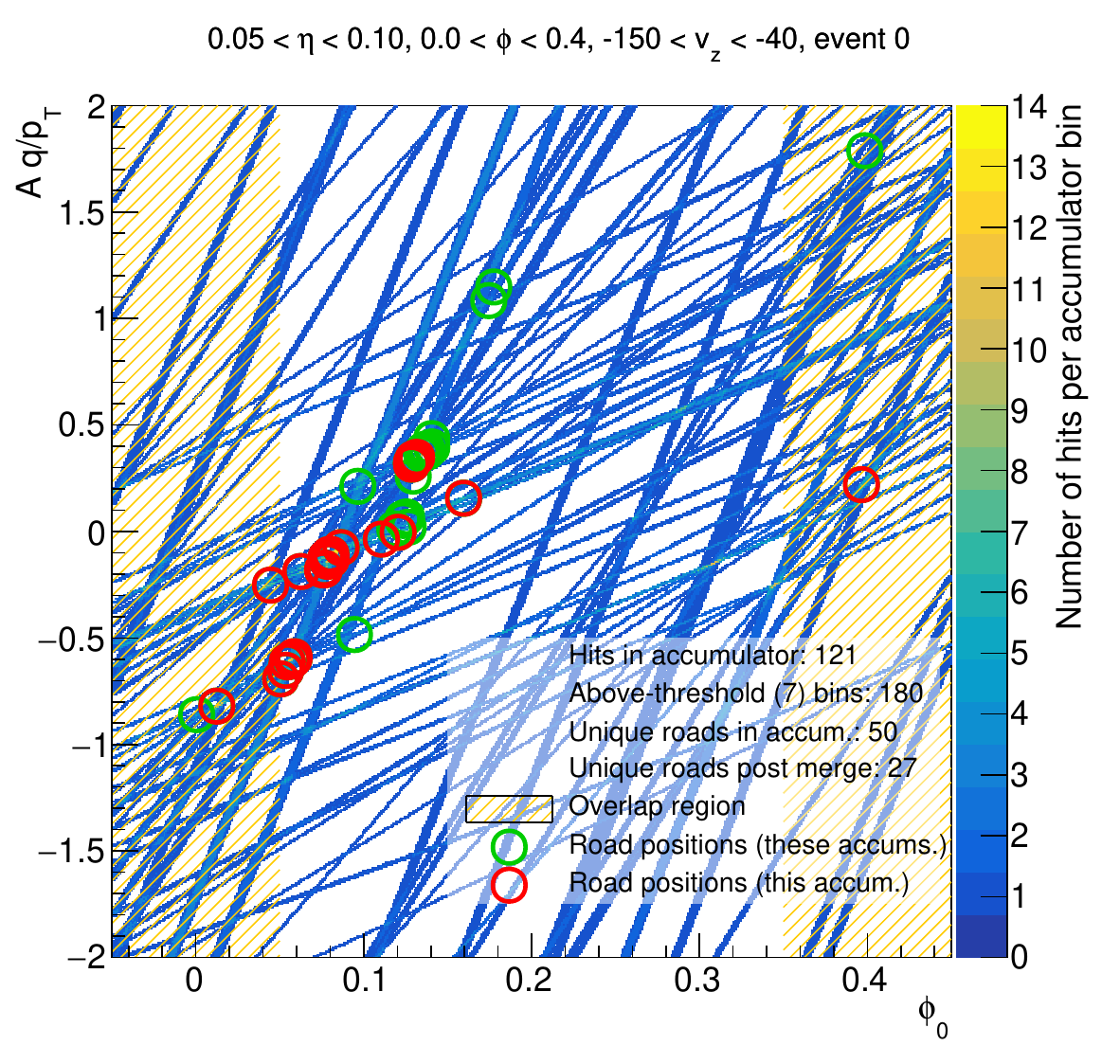}
\includegraphics[width=0.49\textwidth]{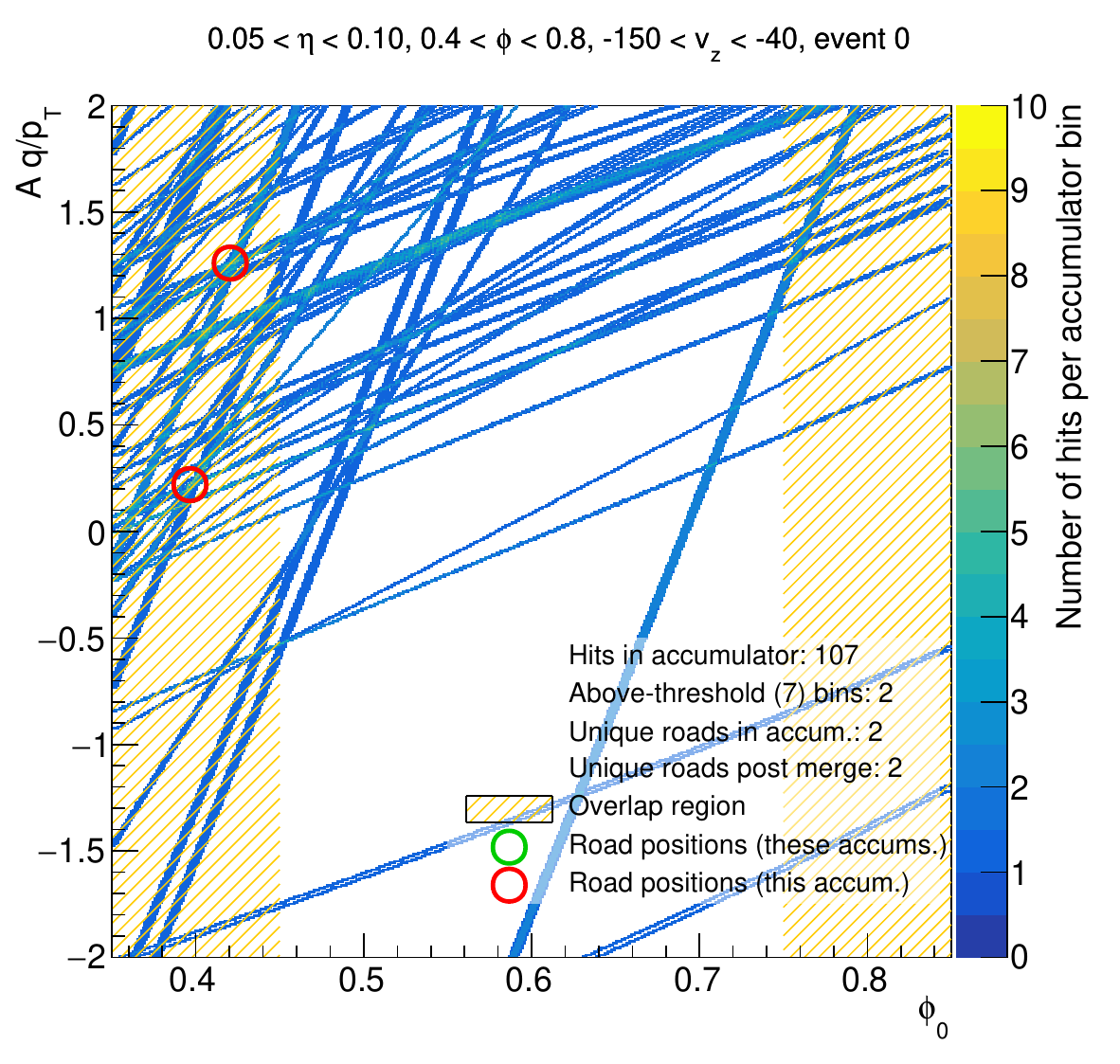}
\includegraphics[width=0.49\textwidth]{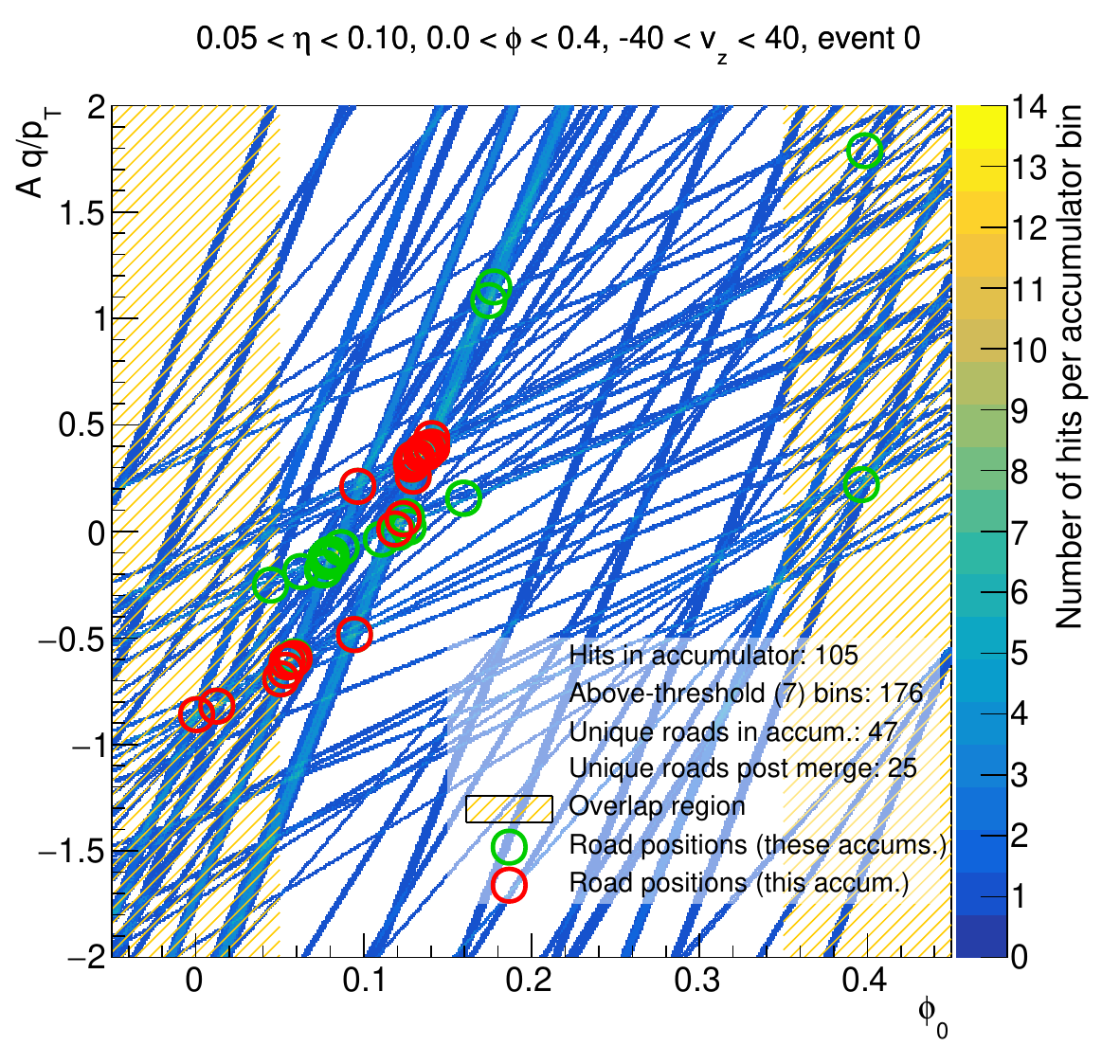}
\includegraphics[width=0.49\textwidth]{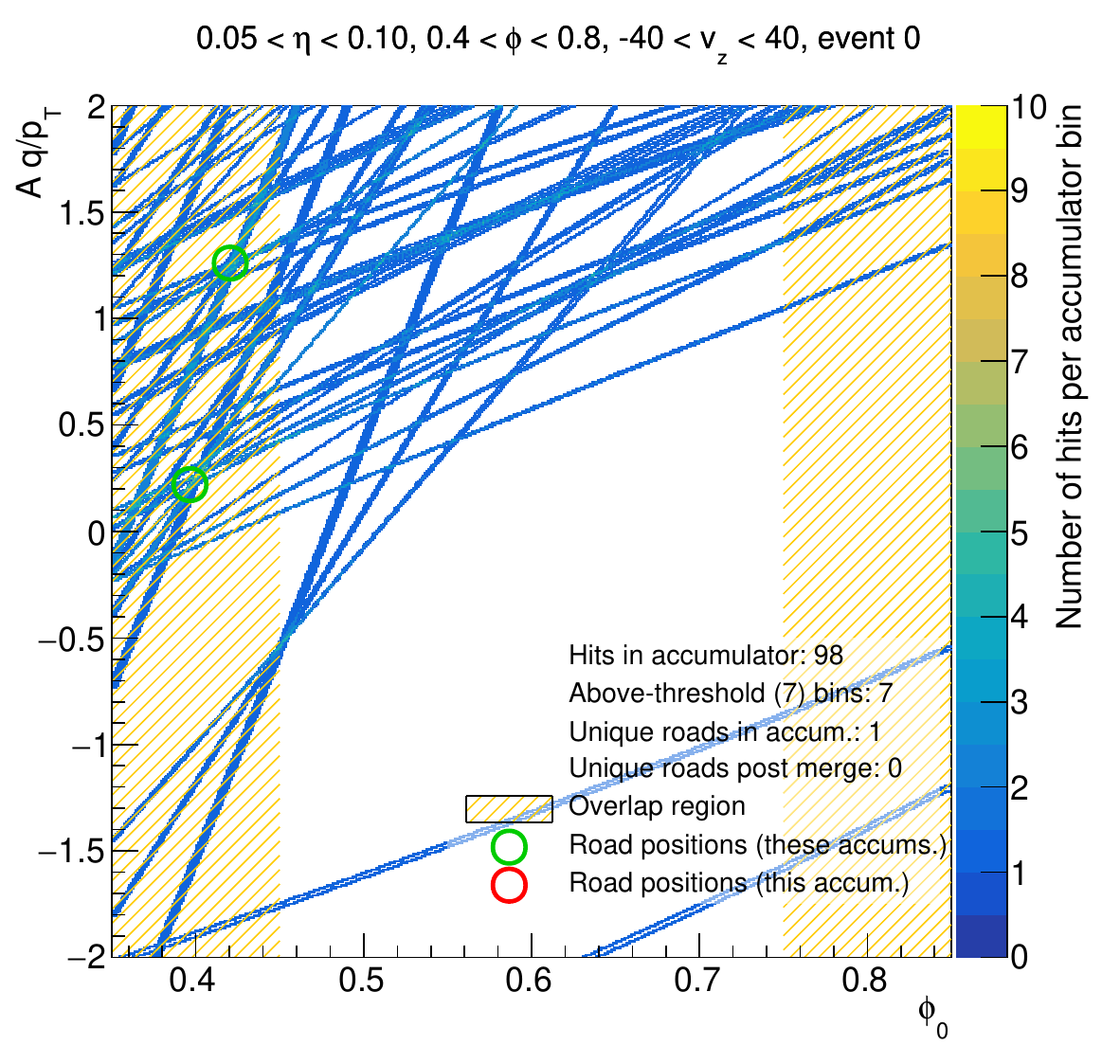}
 \label{fig:accumulators_0}
\end{figure}

In events averaging 16000 hits in this region (|eta|<0.7, extended +- 500mm in $z$) in the relevant detector layers (covering strip, short strip and outer pix (barrel only for all)),
each accumulator has on average around 300 hits thanks to overlaps between them (particularly in the $z/z_0$ slicing) -- ie about 900k accumulator entries total.

This leads to around 650 (1500) bins per accumulator which are above the threshold of 6 (7) -- a collection of hits known as a road. Having removed duplicate and subsets within accumulators (ie, those roads containing a set of hits completely contained in another road), this drops to around 210 (2000) roads per accumulator. After removing overlaps between accumulators of the same eta and phi coordinates but different $z$ slice, this becomes around 75 (500) roads per accumulator. Finally, removing overlaps between overlapping $v_z$ slices leads to about 220,000 (400,000) roads per event, or 75 (140) per accumulator.
The numbers of roads found by this procedure is thus highly dependent on the threshold adopted, which itself has a large impact on the track-finding efficiency.

\begin{figure}
    \centering
\includegraphics[width=0.45\textwidth]{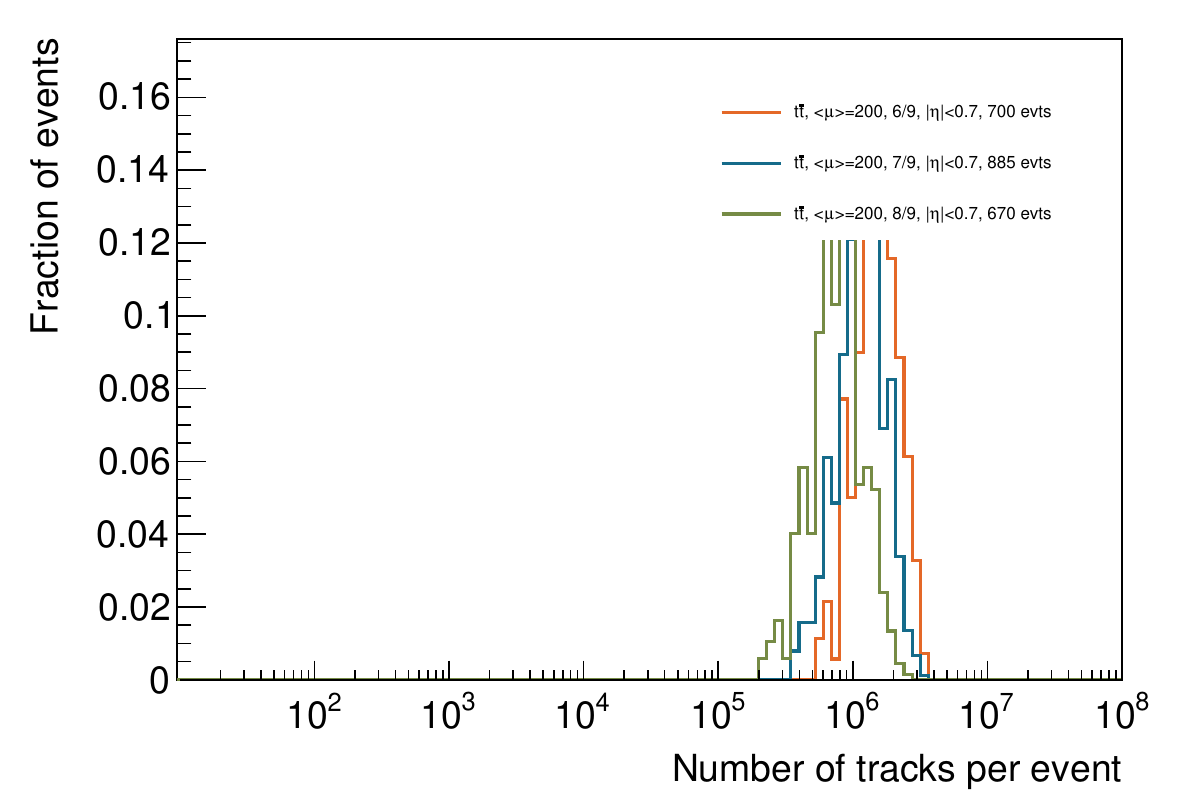} \\
\includegraphics[width=0.45\textwidth]{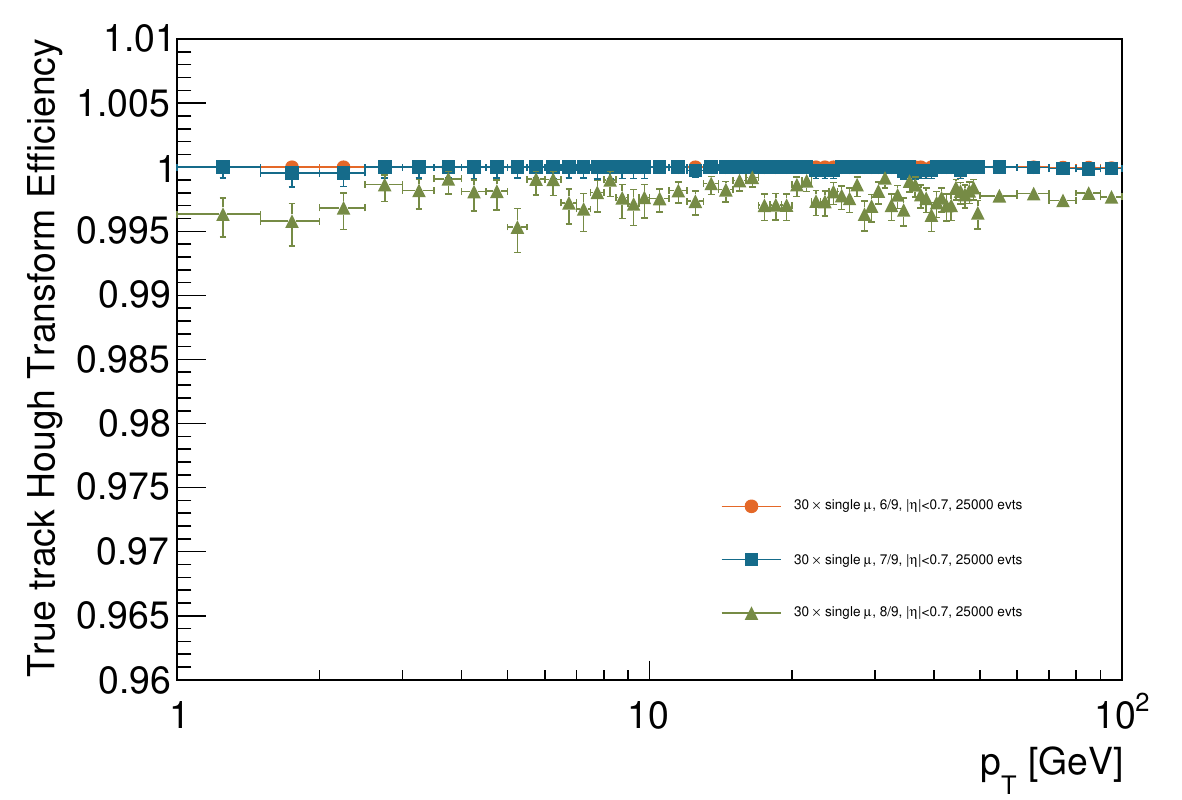}
\includegraphics[width=0.45\textwidth]{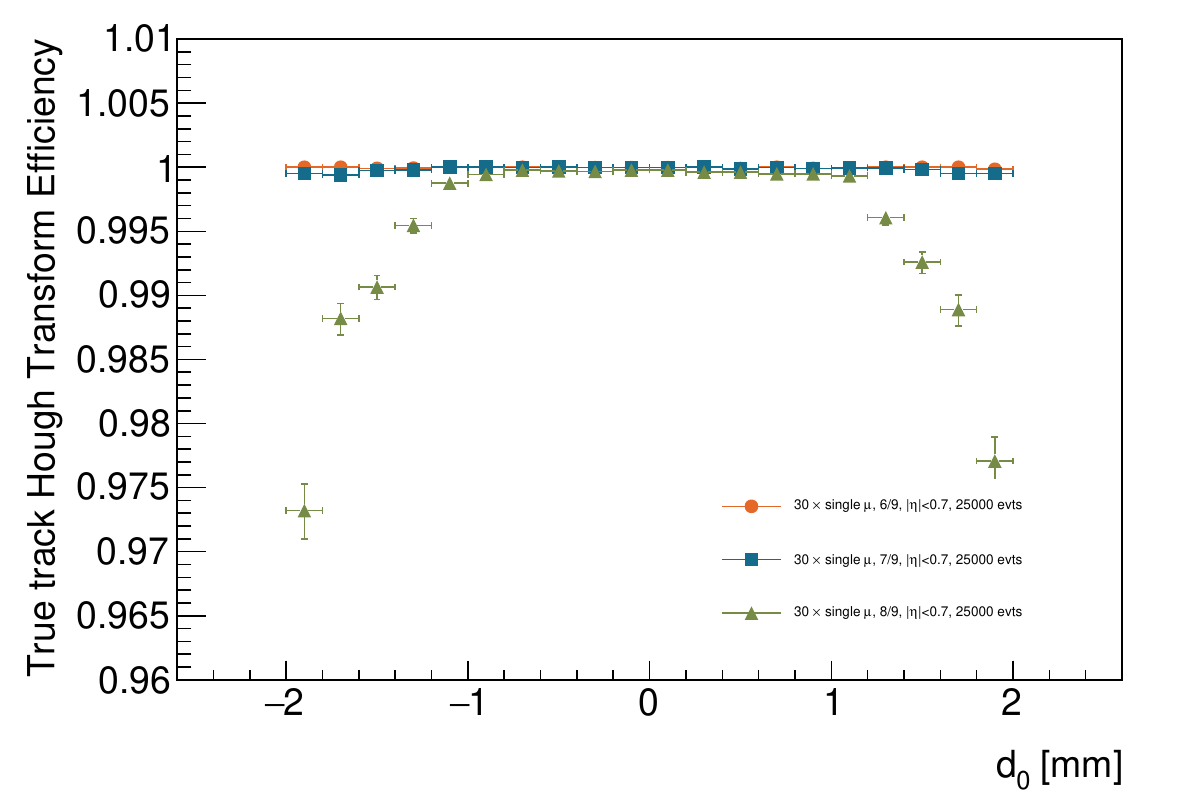}
\caption{Hough Transform performance, considering true tracks within $|\eta|<0.7$ and all $\phi$, for events containing 30 single muons (bottom) and t$\bar{t}$ with 200 pileup events overlaid (top)}
 \label{fig:HTperformance_6v7v8}
\end{figure}

Figure~\ref{fig:HTperformance_6v7v8} shows the evolution of number of track candidates reconstructed and reconstruction efficiency for different accumulator thresholds.

\acknowledgments

This research was supported in part by the U.S. Department of Energy’s Office of Science, Office of High Energy Physics, under Contracts No.  DE-SC0012704, (BNL Laboratory Directed Research and Development Investment LDRD 19-027 and Early Career grant  LAB19-2019, FWP PO201) and DE-SC0011726; the National Science Foundation under Cooperative Agreement OAC-1836650.
This work was supported by resources provided by the Scientific Data and Computing Center (SDCC),
a component of the Computational Science Initiative (CSI) at Brookhaven National Laboratory (BNL).
We would like to thank J. Adelman, E. Brost and S. Majewski  for their help in designing the algorithm for duplicate removal and general discussion.

\bibliographystyle{unsrt}
\bibliography{main.bib}

\begin{thebibliography}{10}

\bibitem{Evans:2008}
{L. Evans and P. Bryant}.
\newblock {LHC Machine}.
\newblock {\em JINST,}, 3:S08001, 2008.

\bibitem{ATLAS:2008xda}
G.~Aad et~al.
\newblock {The ATLAS Experiment at the CERN Large Hadron Collider}.
\newblock {\em JINST}, 3:S08003, 2008.

\bibitem{Apollinari:2017lan}
G.~et~al Apollinari.
\newblock {High-Luminosity Large Hadron Collider (HL-LHC)}: {Technical Design
  Report V. 0.1}.
\newblock 4/2017, 2017.

\bibitem{FCC-hh}
A.~Abada. et~al.
\newblock {FCC-hh: The Hadron Collider}.
\newblock {\em The European Physical Journal Special Topics}, 228, 2019.

\bibitem{CNNS}
Keiron O'Shea and Ryan Nash.
\newblock An introduction to convolutional neural networks, 2015.

\bibitem{RNNS}
Alex Krizhevsky, Ilya Sutskever, and Geoffrey~E Hinton.
\newblock Imagenet classification with deep convolutional neural networks,
  2012.

\bibitem{Tsaris_2018}
Aristeidis~Tsaris et~al.
\newblock The {HEP}.{TrkX} project: Deep learning for particle tracking.
\newblock {\em Journal of Physics: Conference Series}, 1085:042023, sep 2018.

\bibitem{GNNs}
Xiangyang et~al Ju.
\newblock Graph neural networks for particle reconstruction in high energy
  physics detectors, 2020.

\bibitem{Elabd:2021lgo}
Abdelrahman Elabd et~al.
\newblock {Graph Neural Networks for Charged Particle Tracking on FPGAs}.
\newblock {\em Front. Big Data}, 5:828666, 2022.

\bibitem{DeZoort:2021rbj}
Gage DeZoort, Savannah Thais, Javier Duarte, Vesal Razavimaleki, Markus
  Atkinson, Isobel Ojalvo, Mark Neubauer, and Peter Elmer.
\newblock {Charged Particle Tracking via Edge-Classifying Interaction
  Networks}.
\newblock {\em Comput. Softw. Big Sci.}, 5(1):26, 2021.

\bibitem{Ju_2021}
Xiangyang~Ju et~al.
\newblock Performance of a geometric deep learning pipeline for {HL}-{LHC}
  particle tracking.
\newblock {\em The European Physical Journal C}, 81(10), oct 2021.

\bibitem{Aaboud_2017}
ATLAS Collaboration.
\newblock Performance of the {ATLAS} track reconstruction algorithms in dense
  environments in {LHC} run 2.
\newblock {\em The European Physical Journal C}, 77(10), oct 2017.

\bibitem{KalmanFilter}
R~Mankel.
\newblock Pattern recognition and event reconstruction in particle physics
  experiments.
\newblock {\em Reports on Progress in Physics}, 67(4):553--622, mar 2004.

\bibitem{KalmanFilterGPU}
Min-Yu Huang, Shih-Chieh Wei, Bormin Huang, and Yang-Lang Chang.
\newblock Accelerating the kalman filter on a gpu.
\newblock In {\em 2011 IEEE 17th International Conference on Parallel and
  Distributed Systems}, pages 1016--1020, 2011.

\bibitem{NNApproximators}
H.~White K.~Hornik, M.~Stinchcombe.
\newblock Multilayer feedforward networks are universal approximators.
\newblock {\em Neural Networks}, 2:359--356, 1989.

\bibitem{jaeger2007optimization}
Herbert Jaeger, Mantas Luko{\v{s}}evi{\v{c}}ius, Dan Popovici, and Udo Siewert.
\newblock Optimization and applications of echo state networks with
  leaky-integrator neurons.
\newblock {\em Neural networks}, 20(3):335--352, 2007.

\bibitem{HoughTransformDuda}
Richard~O. Duda and Peter~E. Hart.
\newblock Use of the hough transformation to detect lines and curves in
  pictures.
\newblock {\em Commun. ACM}, 15(1):11–15, jan 1972.

\bibitem{HoughTransformOrig}
P.~V. Hough.
\newblock Method and means for recognizing complex patterns.
\newblock {\em US Patent 3,069,654}, 1962.

\bibitem{ATLASTDR}
ATLAS Collaboration.
\newblock Technical design report for the phase-ii upgrade of the atlas trigger
  and data acquisition system - ef tracking amendment, 2022.

\bibitem{ACTS}
Xiaocong et~al Ai.
\newblock A common tracking software project, 2021.

\bibitem{corentin_allaire_2022_6445359}
Corentin Allaire, Paul Gessinger, Julia Hdrinka, Moritz Kiehn, Fabian Kimpel,
  Joana Niermann, Andreas Salzburger, and Stanislava Sevova.
\newblock Opendatadetector, April 2022.

\bibitem{chollet2015keras}
Fran\c{c}ois Chollet et~al.
\newblock Keras.
\newblock \url{https://keras.io}, 2015.

\bibitem{NEURIPS2019_9015}
Adam Paszke, Sam Gross, Francisco Massa, Adam Lerer, James Bradbury, Gregory
  Chanan, Trevor Killeen, Zeming Lin, Natalia Gimelshein, Luca Antiga, Alban
  Desmaison, Andreas Kopf, Edward Yang, Zachary DeVito, Martin Raison, Alykhan
  Tejani, Sasank Chilamkurthy, Benoit Steiner, Lu~Fang, Junjie Bai, and Soumith
  Chintala.
\newblock Pytorch: An imperative style, high-performance deep learning library.
\newblock In H.~Wallach, H.~Larochelle, A.~Beygelzimer, F.~d\textquotesingle
  Alch\'{e}-Buc, E.~Fox, and R.~Garnett, editors, {\em Advances in Neural
  Information Processing Systems 32}, pages 8024--8035. Curran Associates,
  Inc., 2019.

\bibitem{hls4ml}
Farah~Fahim et~al.
\newblock hls4ml: An open-source codesign workflow to empower scientific
  low-power machine learning devices.
\newblock {\em CoRR}, abs/2103.05579, 2021.

\end{thebibliography}

\end{document}